\documentclass[12pt]{iopart}
\expandafter\let\csname equation*\endcsname=\relax 
\expandafter\let\csname endequation*\endcsname=\relax
\expandafter\let\csname leftroot\endcsname=\relax
\expandafter\let\csname uproot\endcsname=\relax
\expandafter\let\csname dddot\endcsname=\relax
\expandafter\let\csname ddddot\endcsname=\relax
\usepackage{amsmath, amssymb}
\usepackage[T1]{fontenc}
\usepackage{todonotes}
\usepackage{graphicx}
\usepackage{dcolumn}
\usepackage{bm}
\usepackage{physics}
\usepackage{ulem}
\usepackage[numbers, square,sort&compress]{natbib}
\usepackage{url}
\usepackage{hyperref}
\usepackage{bbm}
\usepackage{enumitem}

\begin{document}	
\title{Time crystal embodies chimeralike state in periodically driven quantum spin system}
\author{Mahbub Rahaman}
\ead{mrahaman@scholar.buruniv.ac.in}
\address{Department of Physics, The University of Burdwan, Burdwan 713104, India}	
\author{Akitada Sakurai}
\address{Quantum Information Science and Technology Unit, Okinawa Institute of Science and Technology Graduate University, Onna-son, Okinawa 904-0495, Japan}		
\author{Analabha Roy}
\ead{daneel@utexas.edu}
\address{Department of Physics, The University of Burdwan, Burdwan 713104, India}

\begin{abstract}
	Chimera states are a captivating occurrence in which a system composed of multiple interconnected elements exhibits a distinctive combination of synchronized and desynchronized behavior. The emergence of these states can be attributed to the complex interdependence between quantum entanglement and the delicate balance of interactions among system constituents. The emergence of discrete-time crystal (DTC) in typical many-body periodically driven systems occurs when there is a breaking of time translation symmetry. Coexisting coupled DTC and a ferromagnetic dynamically many-body localized (DMBL) phase at distinct regions have been investigated under the controlled spin rotational error of a disorder-free spin-1/2 chain for different types of spin-spin interactions. We contribute a novel approach for the emergence of the DTC-DMBL-chimeralike state, which is robust against external static fields in a periodically driven quantum many-body system.
\end{abstract}
\maketitle

\section{\label{sec:intro} Introduction}
The phenomenon of a \textit{ chimera state} is observed in coupled systems of  identical nonlinear oscillators, when spontaneous synchronized and desynchronized dynamics coexist simultaneously~\cite{kuramoto_coexistence_2002, panaggio_chimera_2015}. Kuramoto et al. first detected this phenomena in a network of non-locally coupled phase oscillators in 2002. Two domains of coherent oscillations with unique frequencies and incoherent oscillations with distributed frequencies were observed.~\cite{kuramoto_coexistence_2002}. These patterns were called ‘chimera states’ by Strogatz~\cite{chimera:strogatz}. Chimeras have been widely explored in classical systems over the last decade~\cite{parastesh_chimeras_2021,chimera_book, taniya2022}. The origin of the chimera lay in the symmetry-breaking bifurcation in the Kuramoto model, which led to a breakdown of global synchronization. This gave rise to a chimera state where spatially distinct regions exhibit different synchronization behaviors~\cite{Kotwal2017}. The coexistence of synchronized and desynchronized states in a chimera state can be considered a manifestation of spontaneous symmetry breaking in the context of nonlinear dynamics~\cite{Aneta2013}. In the physical realm, chimera states serve as a possible explanation of Unihemispheric Slow Wave Sleep (UHSW) in migrating birds, seals and domestic chicks \cite{Rattenborg2000, Rattenborg2006, Rattenborg2016}. 
Chimeras have also been observed in models of electrical power grids, where a synchronous state can be stabilized by tuning the parameters of the generator~\cite{Deng_2024, Motter2013}. 

Eventually, chimera states were realized in the quantum regime as an ordered phase of matter~\cite{bastidas_quantum_2015}. However, it has been difficult to extend classical chimeras, which are heavily reliant on nonlinear dynamics, to purely quantum systems with linear unitary dynamics. As a result, quantum chimeras have had to be described in the semi-classical realm. Nonetheless, the possibility of chimeras in closed quantum systems remains, although one needs to take a different approach to create a quantum system where two different dynamics coexist. In fact, states in which two different dynamics coexist in the same quantum system have already been proposed and reported~\cite{Bastidas2018, Zha2020, sakurai_phys_nodate}.

Interest in the formation of chimeras in magnetic systems has recently increased. Curie-Weiss-type models, such as the Ising model~\cite{singh_chimera_2011}, are used to represent systems of interacting quantum spins where order and disorder coexist. Sakurai \textit{et al.}~\cite{sakurai_phys_nodate} reported the formation of a stable chimeralike state in a one-dimensional spin-$1/2$ chain, which was achieved by surrounding a \textit{Discrete Time Crystal} (DTC) phase with another dynamical state, which is many-body localized (MBL). DTC is a novel state of matter that arises from the breaking of the discrete-time translational symmetry~\cite{else_floquet_2016}. The time crystal (TC) was first proposed by Frank Wilczek in 2012~\cite{wilczek_quantum_2012}, although it was later shown that such states cannot exist as an equilibrium ground state~\cite{Bruno_comment_1, Bruno2013, watanabe_absence_2015}. However, the possibility remains that TC can be realized in a non-equilibrium system, such as Floquet systems. This has been experimentally demonstrated in several many-body systems~\cite{huang2018,taheri_all-optical_2022, Soham2018, zhang_observation_2017, yao_time_2018,frey_realization_2022, rovny_observation_2018, sacha_time_nodate,golletz_basis_2022}. Additionally, TC can occur in prethermal regimes at quasi-infinite temperatures ~\cite{Stasiuk2023} and dissipative systems caused by controlled cavity-dissipation, interactions, and external forcing in experimental settings~\cite{Hans2021}. Recently, a time crystal comb in open systems has been proposed for a dissipative strong interacting Rydberg gas~\cite{jiao2024}. In addition, a continuous time crystal (CTC) where continuous time translational symmetry is broken spontaneously,  has been investigated by introducing an asymmetric subspace in a driven dissipative spin model~\cite{solanki2024}.

Periodic dynamics are typical in quantum mechanical dynamical systems. Well-known examples include Rabi oscillations~\cite{Sakurai_Napolitano_2020}, Zitterbewegung~\cite{LeBlanc_2013}, \textit{etc}. When a quantum system described by a state $\ket{\psi}$ undergoes evolution monitored by the propagator $e^{i H t / \hbar}$~\cite{Biao2018,russomanno_floquet_2017}, it can display distinct oscillations. Therefore, it is crucial to establish criteria that distinguish between well-established periodic events in quantum mechanics and the emergent complex time crystal. The standard definition of a $\mathbb{Z}_2$ symmetry-breaking time crystal is that there exists an observable $\hat{O}$ (corresponding to the order parameter) associated with $\ket{\psi}$, at thermodynamic limit and with the system undergoing oscillation far from equilibrium, the quantity
$\displaystyle f(t) \equiv \lim_{\mathrm{N}\to\infty}\expval{\hat{O}(t)}{\psi}$,
satisfies the following conditions;

\begin{enumerate}[label=(\alph*)]

\item 
Time translation symmetry breaking (TTSB): $f(t) \neq f(t+\tau)$ while $\hat{H}(t + \tau) =  \hat{H}(t)$, 

\item
Rigidity: $f(t)$ exhibits an oscillation of fixed period $T (T=2\tau)$ without requiring any fine-tuning of the  Hamiltonian parameters,

\item
 Persistence: The nontrivial periodicity with a fixed time period $ T $ endures for a sufficient duration to maintain periodicity at infinity as the thermodynamic limit, $\mathrm{N} \rightarrow \infty$ is approached. 
\end{enumerate}
  Thus, the dynamics of a TC should produce a prominent peak in the spectral decomposition of $f(t)$, given by $f_\Omega$, at the TTSB frequency $\Omega=\omega_B \equiv 2\pi/\tau_B$. This manifests in the time dynamics through subharmonic modes, oscillations at integer fractions of the fundamental frequency~\cite{Biao2018, russomanno_floquet_2017}. However, the subharmonic modes are usually unstable, melting the time crystal to the trivial infinite temperature state as predicted by the Floquet Eigenstate Thermalization Hypothesis~\cite{Mori_2018, Kim_2014, Mizuta_2020, Mori_2023_1}. To stabilize the subharmonic modes, a widely practiced method is to introduce  disorder in the model, thus manifesting MBL in the system~\cite{zhang_observation_2017, choi_observation_2017, Khemani2016,else_discrete_2020} The MBL phase tries to preserve the initial state, ideally for an indefinite period of time. This prevents the DTC from melting and thermalizing~\cite{zhang_observation_2017,alet_many-body_2018,else_floquet_2016,smith_many-body_2016,nguyen_signature_2021}, thus allowing the coexistence of two different phases, much like a chimera. It is now of interest to explore whether this chimera can exist in more general cases, such as disorder-free systems where the interactions would not normally lead to localization, but rather to thermalization\footnote{For instance, the Heisenberg interaction has off-diagonal elements, which could potentially disrupt the chimera state by allowing magnetization to travel.}.

In this paper, we explore a novel chimera consisting of a DTC in a disorder-free spin-$1/2$ bipartite chain with Heisenberg exchange interactions. Instead of conventional MBL, we consider \textit{dynamical many-body localization} (DMBL)~\cite{Keser2016, haldar_dynamical_2017,haldar_dynamical_2021,bhattacharyya_transverse_2012,aditya2023dynamical,dutta2014,das_exotic_2010} through \textit{coherent destruction of tunneling} (CDT/DL)~\cite{Grossmann1991,Kayanuma2008} by applying an external time-periodic drive that prevents thermalization in the system. The implementation of a regional spin-flip operation, followed by a global DMBL in a periodic manner, results in the manifestation of TTSB. This leads to the emergence of a DTC phase in one region and DMBL in another region of the spin chain concurrently, a chimera state.  The external periodic transverse drive breaks $\mathbb{Z}_2$ symmetry in the spins in a manner similar to the original proposal~\cite{sakurai_phys_nodate}. We demonstrate that the thermalizing effects of the Heisenberg interaction are suppressed in the chimera state. We also show that the DTC state of the chimera is resilient to additional external static driving fields. 

However, this chimeralike state is distinct from the classical chimera that occurs in systems of identical oscillators. The former is built with non-homogeneous Hamiltonians, contradicting the basic criterion for generic classical chimeras built with identical oscillators under similar environments. Rather, our proposed system is analogous to classical chimeralike nonlocally coupled oscillators~\cite{Jyoti2021,nkomo_chimera_2016}. In addition, quantum mechanics follows the linear Schr\"{o}dinger equation, while classical dynamics in chimeras have to be non-linear. Therefore, one should not expect emergent behavior in purely quantum systems that are identical to those in classical nonlinear systems. The origin of our proposed model is solely quantum mechanical and thus can be labeled as `chimeralike' following the previous works~\cite{sakurai_phys_nodate, Jyoti2021}. 

We present our work as follows: In section~\ref{sec:mdl_n_dynam}, we describe the proposed spin model. In section~\ref{sec:level2}, we describe the emergence of DMBL. In section \ref{sec:level3}, we numerically investigate the coexistence of time crystal and DMBL phases, and support our results analytically. In section \ref{sec:level4}, we look at long-time stability via regional magnetization and entanglement entropy for the system, and explore the robustness of this chimeralike state against external static fields. Finally, we discuss our results and conclude.	
	
\section{\label{sec:mdl_n_dynam} The Model and System Dynamics}
We consider a one-dimensional spin-1/2 chain with N sites.  We modulate the spin chain in time by two repeating sequences of pulse waves, both with the same time period $T$.  The first sequence has a pulse width $T_1$ and modulates the spin chain by transverse fields acting on the $x$ spin axis. The second sequence has pulse width $T_2=T-T_1$, and modulates the spin chain with a spin-spin Heisenberg interaction term along the y-axis and a time-periodic drive in the transverse field for in the $z$ spin axis. Thus, the full Hamiltonian is time-periodic with a period $T=T_1+T_2$, and is given by,
\begin{align}
    \hat{H}(t) = 
    \begin{cases}
        \hat{H_1} , & 0\leq t < T_1,\\
        \hat{H_2} , & T_1\leq t < T,
    \end{cases}
    \label{eq:cleanham}
\end{align}
where,
\begin{align}
    \hat{H_1} = & \hbar g (1-\epsilon_A) \sum_{\substack{\\i \in A}}\hat{\sigma}^x_i + \hbar g (1-\epsilon_B) \sum_{i \in B}\hat{\sigma}^x_i+ \hbar\hat{V}(\hat{\sigma}^{\gamma}),\label{eq:sysham1}\\
    \hat{H_2}(t) = & \hbar\sum_{ij} J_{ij} \hat{\sigma}^y_i \hat{\sigma}^y_{j} +  \hbar h_D \sum_i \hat{\sigma}^z_i + \hbar\hat{V}(\hat{\sigma}^{\gamma}),
    \label{eq:sysham2}
\end{align}
\begin{figure}
    \begin{center}
        \includegraphics[width=10cm]{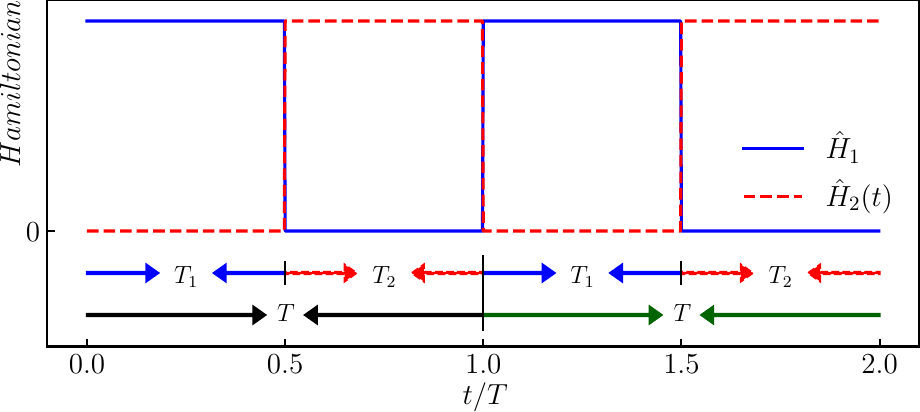}
    \end{center}
    \caption{Pictorial realization of the temporal progression of the Hamiltonian in equation~\eqref{eq:cleanham}. The modulating pulses are set in such a way that the spin-flipping Hamiltonian $\hat{H_1}$, depicted as equation~\eqref{eq:sysham1}, acts during the $T_1-$cycle (blue curve), and the harmonic drive as well as the interactions, depicted as $\hat{H_2}(t)$ in equation~\eqref{eq:sysham2}, acts during the $T_2-$cycle(red dashed curve).}
    \label{Fig:time_distribution}
\end{figure}	
and $\hat{\sigma}^{\mu=x,y,z}_i$ are the Pauli matrices at $i$-th site.  Henceforth, we shall simplify our analysis by assuming that the pulse waves modulating these Hamiltonians have a $50 \%$ duty-cycle, \textit{i.e.} $T_1=T_2=T/2$, as illustrated in figure~\ref{Fig:time_distribution}.  The Hamiltonian $\hat{H}_1$ in equation~\eqref{eq:sysham1} represents a transverse field that can perform spin-flips on spins. To realize the DTC phase on the chain, we set an \textit{ideal value} of this field at $g=\pi/T$. Now, we divide the chain into two physical regions denoted by $A$ and $B$, based on two relative deviations $\epsilon_{A/B}\in[0,1]$ of the field from this ideal value. Thus, these deviation parameters are \textit{rotational errors} for the two regions, and play an essential role in differentiating between the two regions~\footnote{For instance, when $\epsilon_A \sim 0$ and $\epsilon_B \sim 1$, the field in region-A cause imperfect spin-flips, while that in region-B do not flip most spins.}.

The Hamiltonian $\hat{H}_2$ in equation\eqref{eq:sysham2} represents a standard long-range spin-chain in $1-$dimension. The first term models the interaction between two sites $(i>j)$ with a coupling strength $J_{ij}$, assumed to follow a power-law decay $J_{ij}={J_0}/{|i-j|^\beta}$. We have adapted Buyskikh's benchmarking ~\cite{buyskikh_entanglement_2016} to classify the scaling of the spin-spin interaction by varying $\beta$. According to this classification, when $\beta\in\left[0,1\right)$, the interaction is classified as \textit{long-range}, with $\beta=0$ called the \textit{all-to-all} interaction; the range $\beta\in \left(1,2\right)$, as the \textit{ intermediate range} interaction; the range $\beta > 2$ as the \textit{ short-range} interaction, with $\beta= \infty$ called the \textit{ nearest-neighbor} interaction. The second term in $\hat{H}_2$ is a continuous transverse time-periodic drive $\displaystyle \hat{H}_D=\hbar h_D \sum_i\hat{\sigma}^z_i$, where $\displaystyle h_D = -h\sin{(\omega t)}$, and $h,\omega$ are the amplitude and frequency (respectively) of the drive. We have considered $\omega$ to be high enough such that $\omega\gg J_0$. 	Experimentally, this can be accomplished using external high-frequency drives whose time scales are significantly shorter than the relaxation rate due to the interactions in $\hat{H}_2$.~\cite{choi_observation_2017,zhang_observation_2017,Cirac_1995,Blatt_2012}. At specific values of $h$ and $\omega$, our model shows a dynamical state that facilitates the manifestation of CDT/DL, which prevents the system from rapid thermalization. We will elaborate on this in section~\ref{sec:level2}. The final term in both $\hat{H}_{1,2}$, in equations~\eqref{eq:sysham1}, and~\eqref{eq:sysham2}, is $\displaystyle \hat{V}(\hat{\sigma}^{\gamma}) \equiv\gamma  \sum_{i=1}^{N} (\hat{\sigma}^x_i + \hat{\sigma}^y_i)$. It serves as a controlled additional static field that operates on the entire spin chain in the $x$ and $y$ spin axes, respectively.  This term can also be interpreted as a static magnetic imperfection that is modulated by a parameter $\gamma$. We have chosen to ignore this by default, and reintroduce it in section \ref{sec:level4} in order to investigate the robustness of the chimeralike state against such imperfections.
\begin{figure}[t!]
    \centering
    \includegraphics[width=7.cm]{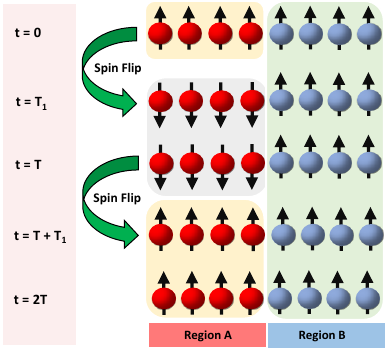}
    \caption{Spin-flips that arise in the spin-$1/2$ chain due to the dynamics described in equations (~\ref{eq:sysham1}) and (~\ref{eq:sysham2}), where the system is broken into two regions, $A$ (left panels) and $B$ (right panels). Snapshots at times $t=0, T_1, T, T+T_1 $ and $t=2T$ are represented from the top to bottom panels, respectively. The spins in the region $B$ remain unchanged for all time. In the region $A$, the spins are flipped in $T_1$ and preserved by CDT/DL during the $T_2-$cycle until $t=T_1+T_2=T$. In the next $T_1-$cycle, they are flipped back and preserved again by CDT/DL until $t=2T$. Thus, the region $A$ has a \textit{period-doubling} response from $t=0$ to $t=2T$.}
    \label{Fig:spinflip}
\end{figure}
	
	We populate the spins in a fully polarized product state of up-spins, as depicted in the top panel of figure~\ref{Fig:spinflip}. During the $T_1-$cycle of the pulses,  the spins in region A \textit{ideally} undergo a \textit{spin-flip} resulting in a spin-down orientation, while the spins situated within region B, remain ideally unaltered. If the dynamics in the $T_2-$cycle of the pulses remains localized by CDT/DL, this state of affairs continues to time $t=T$, as depicted in the next lower panel of figure~\ref{Fig:spinflip}. The localization ensures that each of the spins of the system is independent of each other, preventing any growth of correlations or entanglement between them during this cycle. Now, when the $T_1-$cycle repeats, the spins in region A are flipped back, thereby restoring the spin chain to its initial spin orientation. The dynamics remains localized during the next $T_2-$cycle due to CDT/DL. Thus, when the system is driven up to the next time period $2T$, a spin magnetization \textit{period doubling}~\cite{rovny_31mathrmp_2018, Pan2020}, or \textit{half frequency subharmonic} response is expected. 
	
\section{\label{sec:level2} Interacting Dynamical Localization}

Before investigating the chimeralike state in our model, we explain how the Hamiltonian $\hat{H}_2$ in equation~\eqref{eq:sysham2} exhibits CDT/DL by analytically solving for the dynamics during all $T_2-$cycles. During these times,  the entire spin system is driven by a sinusoidal transverse periodic drive $\hat{H}_D$.
	
Let us employ the \textit{moving frame method}~\cite{haldar_dynamical_2021}, where a unitary transformation is performed to the instantaneous rest frame of the rotation representing the drive. The transformation on the state yields $\ket{\psi(t)}_{mov} = \hat{U}^\dagger(t_0,t) \ket{\psi(t)}$, where 
\begin{equation}
    \hat{U}(t_0,t) \equiv \mathcal{T} e^{-\frac{i}{\hbar}\int_{t_0}^{t} dt' \hat{H}_D}.
    \label{eq:rot1}
\end{equation}
During the first $T_2-$cycle, $t \in{\Big[\frac{T}{2}, T \Big]}$. Thus,
\begin{equation}
    \hat{U}\left(T/2,t\right) = \exp \Bigg[-\frac{i}{\hbar}\int_{T/2}^{t+T/2} (-h \sin(\omega t'))dt'\hbar\sum_i\hat{\sigma}^z_i\Bigg]
    = \prod_{i} \exp\Big[-i \hat{\sigma}^z_i\zeta(t))\Big],
\end{equation}
where, $\displaystyle{
    \zeta (t) = h\int_{T/2}^{T/2+t}  \Big[-\sin(\omega t')dt'\Big]=  \frac{h}{\omega}\Big[1-\cos(\omega t)\Big]}$.		
Now, in this `rotating frame', the Hamiltonian transforms to (see~\ref{sec:AppendixA} for details),
\begin{align}
    \hat{H}^{mov}(t) &= \hat{U}^\dagger \hat{H}_2(t) \hat{U}- i \hat{U}^\dagger \partial_t \hat{U},\nonumber\\
    &=\hbar\sum_{ij}J_{ij}\Big\{\hat{\sigma}^y_i\cos{\big[2\zeta(t)\big]}-\hat{\sigma}^x_i\sin{\big[2\zeta(t)\big]}\Big\}\Big\{\hat{\sigma}^y_j\cos{\big[2\zeta(t)\big]}-\hat{\sigma}^x_j\sin{\big[2\zeta(t)\big]}\Big\},
    \label{eq:movham}
\end{align}
where, in the last step, we have employed the Baker-Campbell-Hausdorff formula~\cite{Magnus1954}. Next, we decompose the moving frame Hamiltonian into its Fourier modes and introduce the Rotating Wave Approximation (RWA), which, in the high-frequency limit, approximates the Fourier modes with their coarse-grained values over long times. This allows us to average out all but the zeroth Fourier mode (see~\ref{sec:AppendixA} for details). This yields
\begin{multline}
\hat{H}^{mov}\approx \hat{H}^{_{RWA}} = \frac{\hbar}{2}\sum_{ij} J_{ij} \hat{\sigma}^y_i\hat{\sigma}^y_j\Bigg\{\left[1+\cos(\frac{4h}{\omega})\mathcal{J}_0\left(\frac{4h}{\omega}\right)\right] \\
+ \hat{\sigma}^x_i\hat{\sigma}^x_j \left[1-\cos(\frac{4h}{\omega})\mathcal{J}_0\left(\frac{4h}{\omega}\right)\right]
+ \left(\hat{\sigma}^x_i\hat{\sigma}^y_j+\hat{\sigma}^y_i\hat{\sigma}^x_j\right)\sin(\frac{4h}{\omega})\mathcal{J}_0\left(\frac{4h}{\omega}\right) \Bigg\}.
\label{eq:movham1}
\end{multline}
Now, if the drive parameters $h$ and $\omega$  are engineered in such a way that the ratio ${4h}/{\omega}$ lies at a CDT/DL point, which is given by one of the roots of the zeroth order Bessel function $\mathcal{J}_0\left(\frac{4h}{\omega}\right)$~\footnote{This can be achieved when $\omega \gg J_0$ by ensuring that $h\gg J_0$.}, it is possible to nullify the dynamics of $\hat{H}_2$ during all $T_2-$cycles, thus inducing localization in the system. We have supported this result by numerical simulations, detailed in section~\ref{sec:level42},where, $\hat{\sigma}^z$ is detected as an approximate integral invariant ~\cite{Keser2016,Dodonov1978}. Maintaining the drive parameters at this CDT/DL point, adiabatically changing $\omega$ from small to large values results in a thermal to DMBL phase crossover~\cite{Mahbub2024}. Therefore, the drive frequency that goes with the CDT/DL point is limited to large values in both $T_2$ and total time period T. To preserve DMBL at the CDT/DL point, we applied a large frequency ($\omega =20$) during numerical investigations, large enough for RWA to hold. Furthermore, to ensure consistency, we set the CDT/DL point to the first root of $\mathcal{J}_0\left(\frac{4h}{\omega}\right)$ in all results.

\section{\label{sec:level3}Coexistence of DTC \& DMBL}	
The underlying dynamics of the periodic Hamiltonian of the chimeralike model outlined in section~\ref{sec:mdl_n_dynam} can be understood by employing the Floquet theory (FT). FT allows us to examine the dynamics at the stroboscopic times, namely, at integer multiples of the time period $T$. The effective Floquet Hamiltonian($H^{\mathrm{eff}}$) for the proposed spin-system that undergoes two periods can be written as follows:
\begin{multline}
    H^{\mathrm{eff}} \approx\frac{\hbar}{2} \sum_{l,m\in A}J_{lm}\hat{\sigma}_l^y\hat{\sigma}_m^y +\frac{\hbar \epsilon_A \pi}{4} \sum_{\substack{l,m\in A\\l\neq m}} J_{lm}\hat{\sigma}^z_l\hat{\sigma}^y_m + \frac{\hbar}{2}\sum_{l,m\in B}J_{lm}\hat{\sigma}_l^y \hat{\sigma}_m^y + \frac{h\hbar}{\pi}\sum_{m \in B}\hat{\sigma}^z_m \\ -\frac{\hbar \pi \epsilon_A}{4T}\sum_{l\in A}\Bigg\{\hat{\sigma}^x_l \bigg[\cos(\hat{\theta}_l)\cos(\frac{4h}{\omega})+1 \bigg] + \hat{\sigma}^y_l \cos(\hat{\theta}_l)\sin(\frac{4h}{\omega})-\hat{\sigma}^z_l \sin(\hat{\theta}_l)\Bigg\}.
    \label{eq:floq_eff3}
\end{multline}
Here, $\displaystyle \hat{\theta}_l \equiv 2 \Big(\sum_{m \in B}J_{lm}\hat{\sigma}^y_m \frac{T}{2} \Big)$ denotes a rotation acting on $\hat{\sigma}^y_l$, the local $y-$spin\footnote{See~\ref{sec:AppendixB} for details.}. 
To examine the influence of rotational error on the chimeralike state, let us begin by considering the scenario where $\epsilon_A=0$. In this situation, the effective Hamiltonian $H^\mathrm{eff}$ in equation~\eqref{eq:floq_eff3}  decouples into terms that live separately in regions A and B, \textit{i.e.},
\begin{equation}
    H^{\mathrm{eff}}_{\epsilon_A=0} =  \frac{\hbar}{2}\Bigg( \sum_{l,m\in A} J_{lm} \hat{\sigma}^y_l\hat{\sigma}^y_m +\sum_{l,m\in B} J_{lm} \hat{\sigma}^y_l\hat{\sigma}^y_m\\+\frac{2h }{\pi}\sum_{m \in B}\hat{\sigma}^z_m\Bigg).
\end{equation}
	
In such a case, the quantum dynamics of the two regions exhibit mutual independence when the system is first prepared in a state that can be factorized as a product of states in the invariant subspaces of regions A and B, respectively. Since we still have a significant degree of flexibility in selecting the driving parameters, it is possible to configure them in such a way as to induce DMBL in the region $B$, while preserving a stable DTC in the region $A$. Let us now increase the parameter $\epsilon_A$ while maintaining a sufficiently high frequency for the drive, so that it is possible to disregard the term $\hat{\theta}_l \approx 0$ due to condition $T<<\hbar/J_0$. Under these circumstances, the strength of the coupling between two regions is significantly influenced by the ratio $4h/\omega$. This dependence is characterized by the amplitudes $(1+\cos(4h/\omega))$ and $\sin(4h/\omega)$, which exhibit non-monotonic behavior. Specifically, when $4h/\omega$ takes the form of $(2n+1)\pi$, where $n$ is a non-negative integer, the decoupling of regions A and B occurs again, regardless of the value of $\epsilon_A$ being nonzero.
\begin{figure}[h!]
    \centering
    \includegraphics[width=10cm]{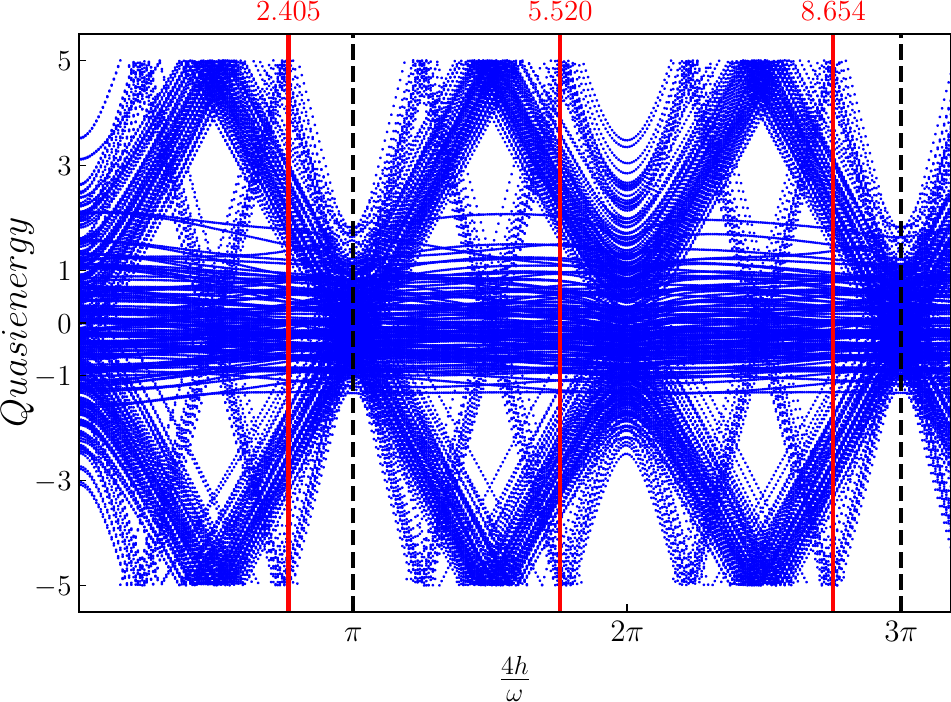}
    \caption{Floquet quasi-energies of the driven spin-chain,  estimated by numerically diagonalizing $H^{\rm eff}$ in equation~\eqref{eq:floq_eff3} at 2T for the one dimensional spin-$1/2$ chain. In this simulation, the number of spins $N=8$ and the strong coupling $J_0=0.2/T$ is considered. The quasi-energies are plotted against $4h/\omega$, where the drive frequency $\omega=20$ is fixed, and drive amplitude $h$ is varied. Level repulsion is minimum when ${4h}/{\omega} = (2n+1)\pi$, $n\in \mathbb{N}_0$. The first two points are shown as a vertical black dashed line. Solid red lines indicate the roots of the zeroth-order Bessel function $\mathcal{J}_0(4h/\omega)$.}
    \label{Fig:quasienergy_new}
\end{figure}	

\begin{figure*}[t!]
    \centering
    \hspace{2cm}\includegraphics[width=13.5cm]{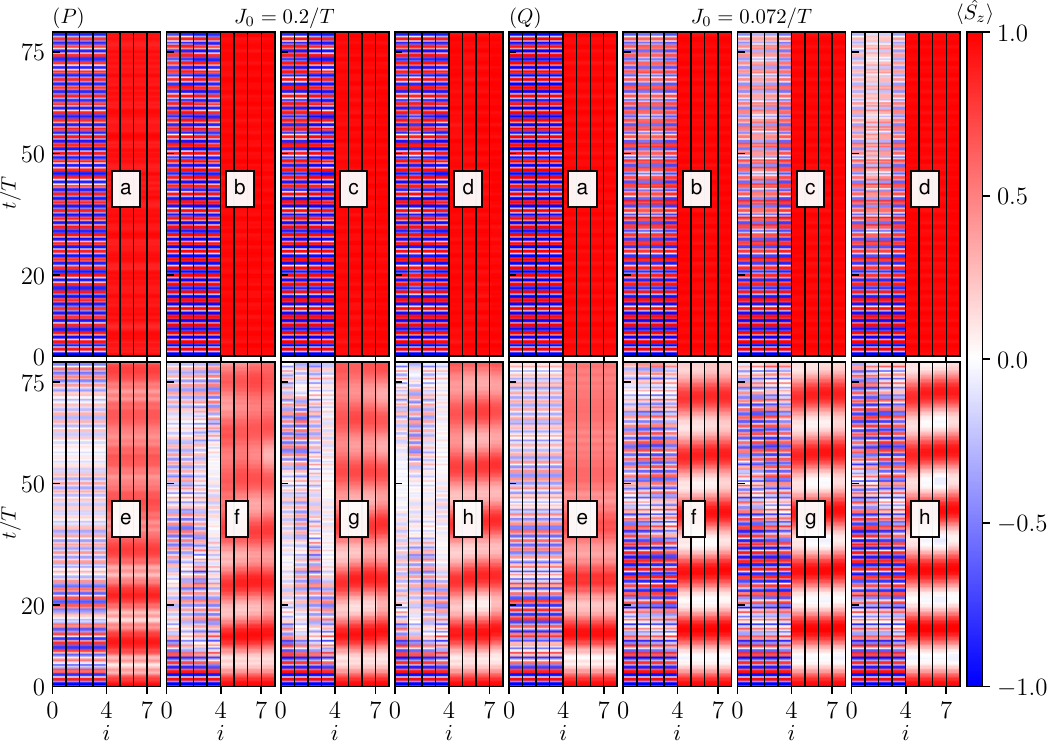}
    \caption{The time evolution of local magnetization for each spin at region A (i=0,1,2,3) and region B (i=4,5,6,7) is plotted for $N=8$ spins in a periodically driven one-dimensional spin chain, starting from a fully polarized spin state, with drive frequency $\omega=20$. The spin rotation error $\epsilon_A = 0.03$ and $\epsilon_B = 0.9$, and $g=\pi/T$, where $T=2\pi/\omega$ is the time period of the drive.  In all two panels, different ranges of spin-interactions such as long-range($\beta=0$), intermediate range($\beta=1.5$), short range($\beta=2.5$), and nearest neighbor range($\beta=\infty$), are chosen and plotted, respectively, from left to right, namely (a, b, c, d) in the top row and (e, f, g, h) in the bottom row of each panel. The results of the time evolution for the strong spin coupling ($J_0 = 0.2/T$) in the left panel (P) and the weak spin coupling ($J_0 = 0.072/T$) are shown in the right panel (Q). The top panels of P \& Q show plots where the drive amplitude $h$ is chosen so that the system lies at the CDT / DL point which is chosen to be the first root of $\mathcal{J}_0\left(\frac{4h}{\omega}\right)$, and the bottom panels show the results away from that CDT/DL point.}
    \label{Fig:strong_weak_ea}
\end{figure*}

We have utilized numerical estimations of eigenvalues to represent the effective Hamiltonian $H^{\rm eff}$. These are connected to the \textit{ Floquet quasienergies} of the driven spin chain, given by the eigenvalues of the time-dependent observable $\hat{H}(t)-i\hbar\pdv*{}{t}$ evaluated at $t=2T$. The numerical diagonalization of the effective Hamiltonian $H^{\rm eff}$ has been performed for a system of $N=8$ spins. The resulting quasi-energies have been displayed as a function of ($4h/\omega$). The results are depicted in figure~\ref{Fig:quasienergy_new}. The quasi-energy distribution displays a consistent pattern that is confined to the Floquet-Brillouin zone, namely the interval of $[-\frac{\pi}{2T}, \frac{\pi}{2T}]$, a consequence of Floquet's Theorem~\cite{dutta2014}. It is evident that the minimum repulsion between the quasi-energies occurs when the parameter $4h/\omega$ is an odd integer multiple of $\pi$. This result offers empirical evidence that is consistent with the theoretical assertions presented earlier in this section on the decoupling of region A and B, even in the cases where $\epsilon_A \neq 0$. Nevertheless, the CDT/DL condition in region-B necessitates that $4h/\omega$ lie on one of the roots of the Bessel function. This condition retains the existing coupling between regions A and B for a nonzero $\epsilon_A$. Therefore, $\epsilon_A$ is the key system parameter to control the coupling between the two regions of our proposed model. 	
	
To gain comprehension of the existing phases in the spin chain, we consider the local magnetization at individual sites (i) as the physical order parameter. We utilize it to investigate the regime where TTSB with a sufficiently rigid DTC occurs, detecting the DTC phase in region A and the DMBL phase in region B. Additionally, the spectral decomposition of the local magnetization can be obtained by Fast Fourier Transforms (FFT) of the time-series data. The FFT can be used to identify the DTC phase by detecting peaks corresponding to the subharmonic modes as defined by the DTC criteria (see section~\ref{sec:intro}). We now present visuals of the local magnetization, denoted as $\expval{\hat{S}^z_i (t)}$, which is equivalent to the expectation value of the operator $\hat{\sigma}^z_i$ with respect to the state $\ket{\Psi(t)}$, for all sites $i$ in the spin-chain. The data mentioned above have been obtained using numerical simulations of Schr\"odinger dynamics, represented by the equation $\hat{H}(t)\ket{\Psi(t)}= i \hbar \pdv*{\ket{\Psi(t)}}{t}$. This equation describes the behavior of the quantum many-body state $\ket{\Psi(t)}$. Simulations were carried out using QuTiP, a Python-based Quantum Toolbox~\cite{Johansson2013}, and the source code is available online~\cite{github_repo}. The simulations were performed over an extended duration (up to $t = 80T$, with $\hbar$ normalized to unity) and over various ranges of power law interactions ($\beta = 0,1.5,2.5,\infty$). Furthermore, simulations were carried out for weak and strong interaction amplitudes ($J_0 = 0.072/T$ and $J_0 = 0.2/T$ respectively), which are similar to the parameters selected by in previous study~\cite{sakurai_phys_nodate,zhang_observation_2017}. To assess the durability of this chimeralike state, we set the rotational error values as $\epsilon_A = 0.03$ and $\epsilon_B = 0.9$. Furthermore, to observe the emergence of the chimeralike state in our model, we manipulate the periodic drive in $\hat{H}_2(t)$ by setting a high frequency $\omega=20$ for two different scenarios: one where the parameter $4h/\omega$ is located at the CDT/DL point, and another where $4h/\omega$  does not lie at any roots of $\mathcal{J}_0$.

As illustrated in figure~\ref{Fig:strong_weak_ea}, the local magnetization exhibits the breaking of discrete time translational symmetry, manifesting a DTC phase in region A at the CDT/DL point, under the condition of strong spin-spin coupling. Simultaneously, region B exhibits localized dynamics. As a result, the coexistence of the DTC and DMBL phases is observed in regions A and B, respectively, leading to the formation of a chimeralike state. Conversely, it is evident that the system dynamics demonstrate instability upon deviation from the CDT/DL point, irrespective of the particular values of coupling strength($J_0$) that have been chosen. In a similar vein, it can be noted that weak interactions demonstrate the presence of a stable chimeralike state when they possess a range that encompasses all spins ($\beta=0$). However, when considering intermediate ($\beta = 1.5$), short ($\beta = 2.5$), and nearest-neighbor range interactions ($\beta = \infty$), the DTC melts rapidly inside region A, typically occurring within $20$ cycles of the drive. This result is consistent with the findings of previously published work~\cite{sakurai_phys_nodate}.
	
The stability of this DTC phase exhibits notable disparities between strong and weak-coupling interactions. At the CDT/DL point, it is seen that the DTC phase exhibits enhanced stability in the presence of strong coupling compared to weak coupling, specifically for interactions that are not long-range in nature. The phenomenon of stable DTC is observed in long-range spin chains, regardless of the strength of spin interactions. Therefore, it is feasible to see the emergence of a chimeralike state consisting of two separate phases of matter, specifically a time crystal and a localized ferromagnet, simultaneously in a spin-chain system. This phenomenon can occur in typical experimental settings, with long-range interactions being the most suitable candidate.

\begin{figure}[h]
	\centering
	\includegraphics[width=10.cm]{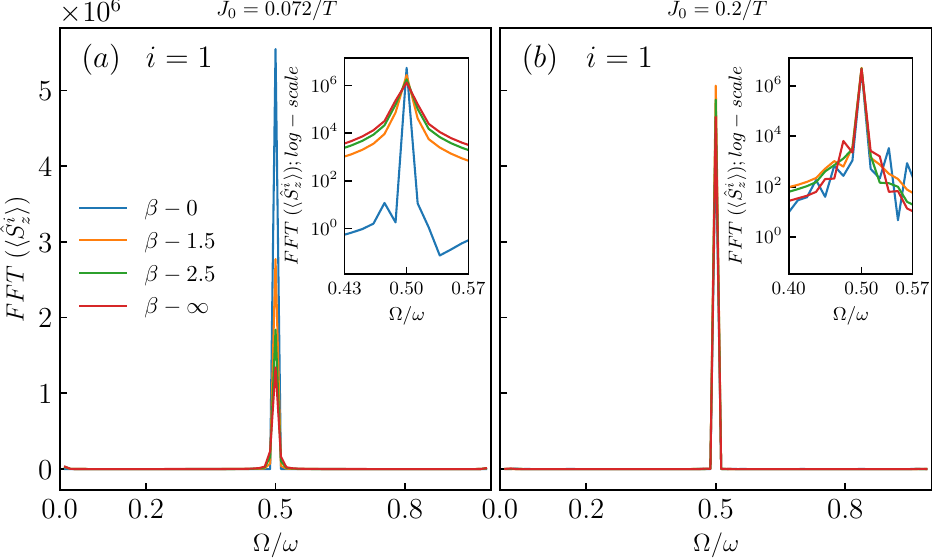}
	\caption{Fast Fourier Transforms(FFT) of the local magnetization $\expval{\hat{S}_z^i}$ at site $i=1$ in region A of the spin-chain for weak coupling ($J_0 = 0.072/T$, panel-a) and strong coupling ($J_0 = 0.2/T$, panel-b) for different values of $\beta$ as indicated in the legend. The frequency $\omega =20$ and amplitude $h$ is adjusted so that the system is at the first root of $\mathcal{J}_0\left(\frac{4h}{\omega}\right)$ \textit{i.e.} CDT/DL point. All other parameters are the same as those in figure~\ref{Fig:strong_weak_ea}. In the inset figure of each of the panels, FFT of $\expval{\hat{S}_z^i}$ is plotted in log-scale for smaller $\Omega/\omega$ range. The peaks at half-integer multiples of $\Omega/\omega$ denote subharmonic responses in region A that result in DTC.}
	\label{Fig:sz_single}
\end{figure}

The suitability of long-range interactions is examined by analyzing the dynamics in frequency space (with the frequency variable denoted by $\Omega$) by applying a Fast Fourier Transform (FFT) on the local magnetism at a specific spin site $(i=1)$ in region A. Numerical FFT algorithms provided by the NumPy library~\cite{harris2020array} were utilized to achieve this task. The results are visually presented in figure~\ref{Fig:sz_single}.  The frequency $\Omega$ in the x-coordinate has been scaled relative to the driving frequency $\omega$. The subharmonic response at $\Omega=\omega/2$ can be detected by observing the peaks at half-integer multiples of $\omega$. This observation serves as confirmation of the onset of discrete TTSB throughout all selected spin-interaction ranges. The stability of the resulting DTC will be influenced by the contributions originating from the underlying continuum of frequencies. When $\beta$ approaches zero, the continuum exhibits a higher degree of submissiveness toward subharmonic peaks compared to higher values. This distinction is especially noticeable under conditions of weak coupling. Therefore, in the context of a DTC phase of matter, it is advantageous to consider long-range interactions rather than short-range interactions when selecting potential candidates. The present analysis is consistent with the numerical results shown in figure~\ref{Fig:strong_weak_ea}, where each individual spin interaction type provides evidence of a reliable DTC.

\begin{figure}[t]
\centering
\hspace{1.5cm}\includegraphics[width=12cm]{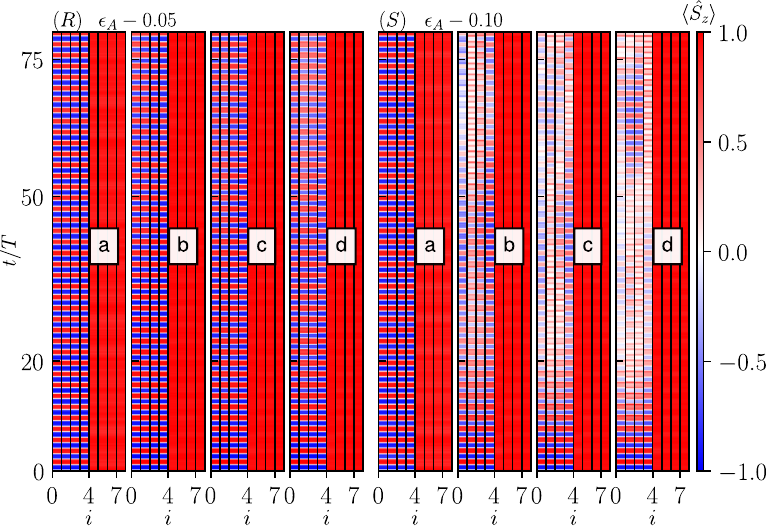}
\caption{The time evolution of local magnetization for each spin at region A (Site(i)=0,1,2,3) and region B (Site(i)=4,5,6,7) is plotted for $N=8$ spins in a periodically driven one-dimensional spin chain, starting from a fully polarized spin state, with drive frequency $\omega=20$. The strong spin-spin coupling $J_0 = 0.2/T$, spin rotation error $\epsilon_B = 0.9$ and $g=\pi/T$, where $T=2\pi/\omega$ is the time period of the drive.  Different ranges of spin-interactions such as long-range($\beta=0$), intermediate range($\beta=1.5$), short range($\beta=2.5$), and nearest neighbor range($\beta=\infty$), are chosen and plotted  for $\epsilon_A =0.05$(R-panel) and $\epsilon_A =0.1$(S-panel), respectively, from left to right, namely (a, b, c, d) in each panel.}
\label{Fig:ea}
\end{figure}
Furthermore, we have conducted an analysis on the resilience of the DTC in the presence of larger rotational errors ($\epsilon_A$). The simulation results are presented in figure~\ref{Fig:ea}. In the present simulations, we examine two different values for the parameter $\epsilon_A$, namely $\epsilon_A = 0.05$ and $\epsilon_A = 0.1$, while maintaining a constant value of $\epsilon_B$ at $0.9$. In this study, we investigate various ranges of interactions, denoted by the parameter $\beta$, which assumes the values of $0.0, 1.5, 2.5$ and infinity. The aforementioned investigations are carried out at a CDT/DL point under strong coupling, $J_0 = 0.2/T$.  A steady DTC was observed for all interaction ranges for simulation times of upto $t=80T$ when the value of $\epsilon_A$ was set to $0.05$. The stability of the DTC is maintained for higher values of $\epsilon_A$, that is, when $\epsilon_A = 0.1$ only when all-to-all interactions ($\beta=0$) in considered. However, in the case of spin-spin interactions with other ranges, the DTC diminishes rapidly. This finding provides further evidence in favor of the proposal that long-range interactions are the most advantageous for the development of a DTC-MBL chimeralike state. This observation holds true for the cases where the spin rotational error $\epsilon_B$ is large ($\sim 0.9$), while $\epsilon_A$ remains relatively small ($\le 0.1$).

\begin{figure}[t!]
	\centering
	\hspace{2cm}\includegraphics[width=13.5cm]{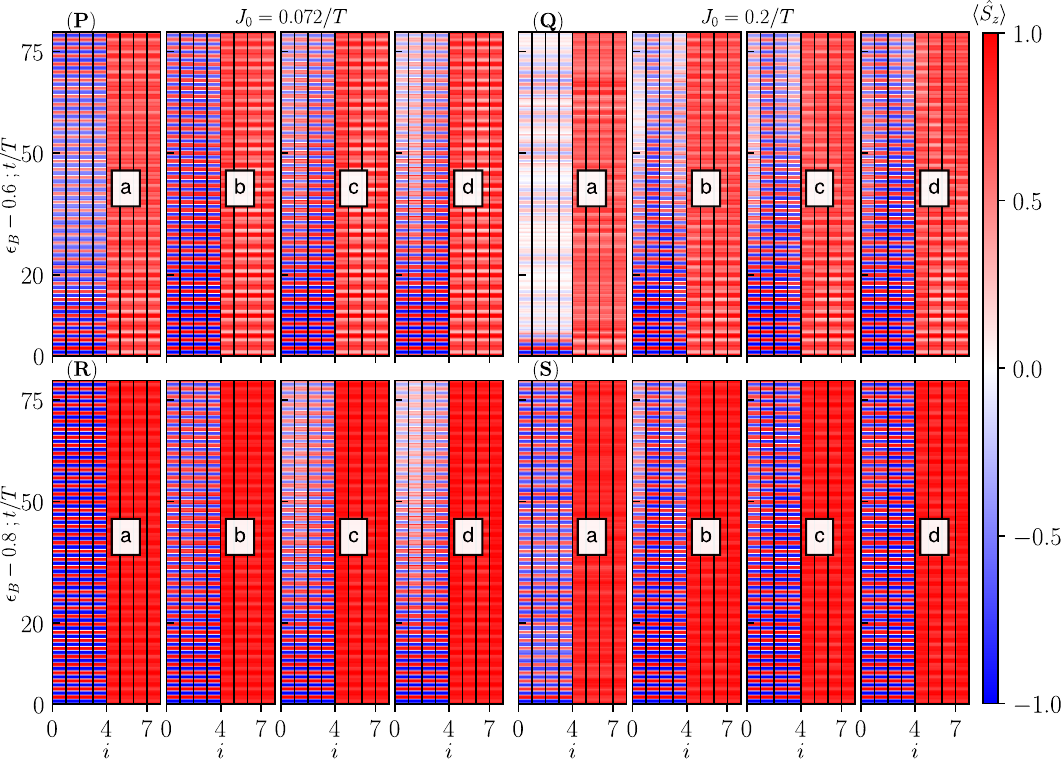}
	\caption{The time evolution of local magnetization for each spin at region A (i=0,1,2,3) and region B (i=4,5,6,7) is plotted for $N=8$ spins in a periodically driven one-dimensional spin-1/2 chain, starting from a fully polarized spin state. We have considered the drive frequency $\omega=20$ and amplitude corresponding to CDT/DL point. The strong spin-spin coupling $J_0 = 0.2/T$ (panels Q \& S) and weak spin-spin coupling $J_0 = 0.072/T$ (panels P \& R),  spin rotation error $\epsilon_A = 0.03$ and $g=\pi/T$, where $T=2\pi/\omega$ is the time period of the drive. The upper panels (P \& Q) denote plots for $\epsilon_B=0.6$ and lower panels (R \& S) denote plots for $\epsilon_B=0.8$. Different ranges of spin interactions such as long-range($\beta=0$), intermediate range($\beta=1.5$), short range($\beta=2.5$), and nearest neighbor range($\beta=\infty$), are chosen respectively, from left to right, namely (a, b, c, d) in each panel.}
	\label{Fig:eb}
\end{figure}
To investigate the impact of $\epsilon_B$ on the emergence of a chimeralike state, we have considered the spin-1/2 chain starting from full z-polarization. We have considered both strong ($J_0 = 0.2/T$) and weak ($J_0 = 0.072/T$) spin-coupling strength, with drive frequency $\omega = 20$, amplitude corresponding to the CDT/DL point, and $\epsilon_A=0.03$. Plots of the local magnetization in time are shown in figure~\ref{Fig:eb} for different spin-spin interaction ranges for two values of $\epsilon_B = 0.6, 0.8$. For both values, the DMBL phase in region B is found to destabilize over different time-scales, despite being at CDT/DL. The instability in DMBL is inversely dependent on $\epsilon_B$. Similarly, the DTC phase in region A eventually melts for all cases of $\epsilon_B$ and interaction ranges, although the melting rate varies depending on the values of $\epsilon_B$ and interaction ranges. For a particular $\epsilon_B$ (say $\epsilon_B=0.8$), each spin in region A exhibits similar dynamics over time only for all-to-all interactions. However, for other shorter interaction ranges, the dynamics shows more spatial variation. The spins in region A that are far away from the bipartite junction melt from DTC faster than the spins closer to the junction. The melting is faster in case of weak coupling in comparison to strong coupling. However, in the case of all-to-all interactions, the DTC-melting behavior is opposite to that of shorter ranges; the DTC phase melts more slowly in weak coupling than the strong coupling. In the case of the smaller $\epsilon_B=0.6$ we have observed that for each of the cases of coupling strength and interaction ranges, the melting is faster than that of $\epsilon_B=0.8$.

All the spins in region A interact equally with all the spins in region B for the all-to-all interaction range ($\beta=0$). This results in equal melting rates of the DTC. For shorter interaction ranges, the spins in region A that are far away from the bipartite junction have weaker interactions with the spins in region B than those that are closer. This results in faster local melting in the DTC phase for the former than the latter. Thus, the local melting-rate of the DTC increases with decreasing interaction range. This corroborates earlier results that indicated better stability of the chimeralike state in all-to-all interactions, as well as the decay of stability with increasing $\beta$. For higher coupling strengths, the amplitude of interaction is larger. This explains why the chimeralike state melts faster in strong coupling and all-to-all interactions than in weaker ones. Additionally, for shorter-range interactions, at strong coupling strength, the chimeralike state melts more slowly compared to weak coupling. Thus, for smaller rotational errors $\epsilon_B$, the DTC-DMBL chimeralike state is not sustainable for long times.

Consequently, we have kept a small value of $\epsilon_A = 0.3$ and a high $\epsilon_B = 0.9$ value to sustain stable chimeralike state. Also, a strong coupling is preferable to investigate the short time dynamics of the DTC-DMBL chimeralike state.

\section{\label{sec:level4} Stability of the chimera phase}
In this section, we investigate the stability of the chimeralike state in various regions of the complex parameter space. We ran simulations for both strong and weak coupling and a wide variety of interaction ranges. The physical phenomena of interest involve the tunneling of any quantum information between the DTC and the DMBL regions that would contribute to melting, and can be profiled by looking at the regional magnetization as well as the entanglement entropy between the regions, as functions of time. The nontrivial control over regional information is a new way to preserve quantum information in non-Markovian open systems, particularly information written to  clean qubits in Noisy Intermediate Scale Quantum (NISQ) processors. Finally, we investigate the robustness of the chimeralike state against external static fields.
\subsection{\label{sec:level42} Regional Magnetization}
\begin{figure}[t]
	\centering
	\includegraphics[width = 15.0cm]{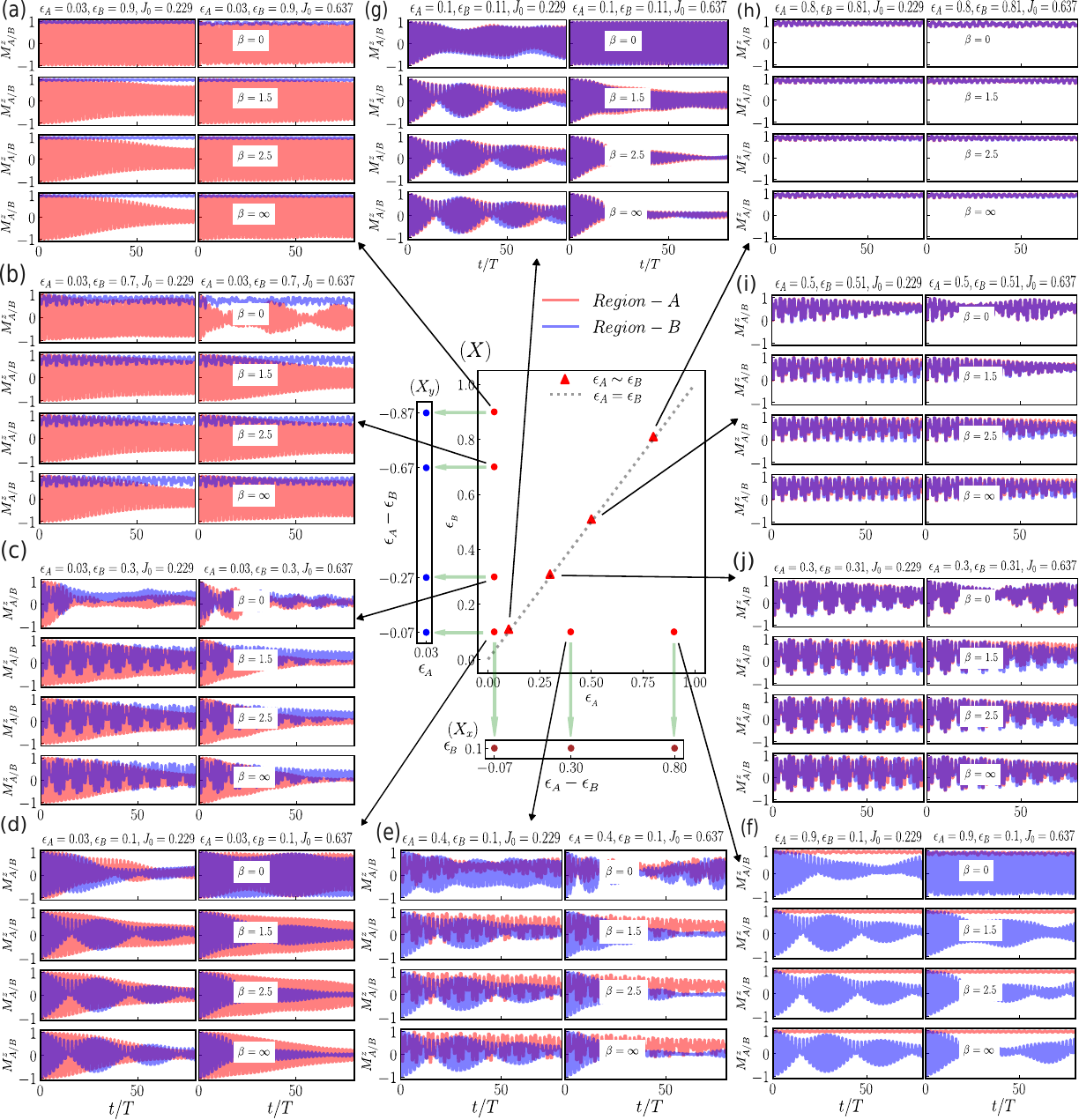}
	\caption{The temporal variation in regional magnetization $M^z_{A/B}$ of the spin-1/2 chain for different spin rotational errors ($\epsilon_{_{A,B}}$) and spin interaction ranges ($\beta$). The simulations were conducted for a fixed time period of 80T, with drive parameters set to a CDT/DL point. The central panel-X illustrates the selection of $\epsilon_{A,B}$ values and connects them to the corresponding plots. The left panels (a-d) show the plots for different values of $\epsilon_B$ (0.1, 0.3, 0.7, 0.9) with a constant $\epsilon_A = 0.03$. The bottom panels (d-f) show the plots for a constant $\epsilon_B = 0.1$ and varying $\epsilon_A$ (0.03, 0.4, 0.9). The right panels (g-j) show the plots for a set of $\epsilon_{A,B}$ values that are comparable ($\epsilon_A\sim \epsilon_B$). Weak ($J_0 = 0.072/T = 0.229$) and strong ($J_0 = 0.2/T = 0.637$) spin coupling strengths are considered. The differences in selected $\epsilon_{A,B}$ values are plotted in the $X_y$ and $X_x$ panels.}
	\label{Fig:reg_mag_ea_eb}
\end{figure}

We now examine the regional magnetization of the chimeralike state, which is given by 
$M^z_{A/B}\equiv\frac{2}{N}\sum_{i\in{A/B}}\expval{\hat{\sigma}^z_i(t)}$, the expectation value  of the total $z-$spin in a specific region. This will strengthen the case for
the suitability of long-range interactions in DTC~\cite{sakurai_phys_nodate}. We have conducted extensive simulations for various spin rotational errors ($\epsilon_{_{A,B}}$) in order to provide a detailed analysis of the stability of the DTC-DMBL chimeralike state. We have explored different values of rotational error by keeping one of the $\epsilon_{_{A,B}}$ constant and adjusting the other from lower to higher values. In addition, we have adjusted both $\epsilon_A$ and $\epsilon_B$ together while keeping them comparable, \textit{i.e.}, $\epsilon_A \sim \epsilon_B$. The simulations yielded the regional magnetization for times up to $80T$. We have  considered various spin-spin interactions, where $\beta = 0,1.5,2.5,$ and $\infty$. All spins are initially z-polarized. The drive parameters ($h,\omega$) are set to a CDT/DL point with $\omega =20$. 
	
The resultant plots can be seen in figure~\ref{Fig:reg_mag_ea_eb}. In this figure, the central primary panel-X shows the selected values of $\epsilon_A$ and $\epsilon_B$. Arrows are drawn in this figure to connect the points in the $\epsilon_A - \epsilon_B$ space to the corresponding $M^z_{A/B}$ plots. Panels 'a' through 'd' show the temporal variation of $M^z_{A/B}$ for different values of $\epsilon_B$ (0.1, 0.3, 0.7, 0.9) with a constant $\epsilon_A = 0.03$. The stability of the DTC-DMBL chimeralike state is observed for all values of $\epsilon_B$ and spin interaction ranges. The DMBL phase remains stable for all spin interaction ranges. However, the DTC phase melts faster as the interaction range shortens. On the other hand, the DTC-melting rate is slower in the case of strong coupling. For smaller values of $\epsilon_B$ (0.7, 0.3, 0.1), the stability of the DMBL phase decreases as $\epsilon_B$ decreases, and the DTC phase melts faster. We have collected data for $M^z_{A/B}$ for comparable values of $\epsilon_{A,B}$ in panels `g', `j', `i', and `h' of figure~\ref{Fig:reg_mag_ea_eb}. As the values of $\epsilon_{A}\sim\epsilon_{B}$ increase, the DTC phase gradually melts and the DMBL phase becomes dominant in both regions A and B.
\begin{figure}[t]
	\centering
	\hspace{1.5cm}\includegraphics[width = 11cm]{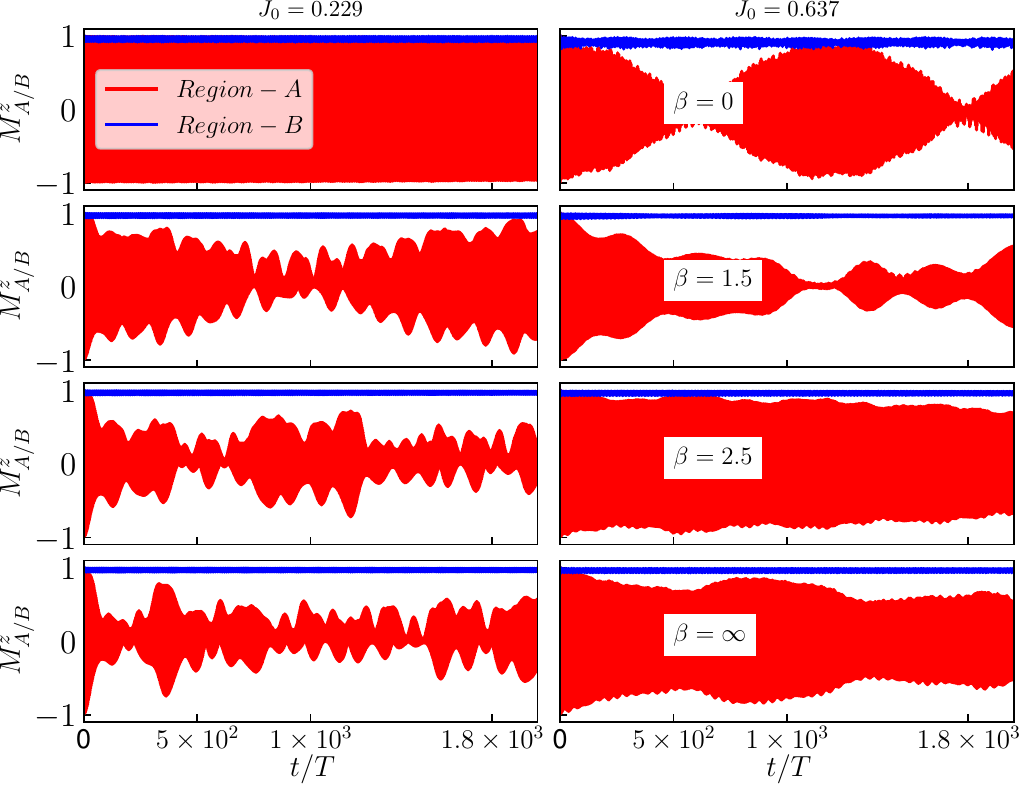}
	\caption{Regional magnetization $M^z_{A/B}$ of the spin-chain. The magnetization of regions A (red) and B (blue) is plotted as functions of time $t/T$ up to 2000T for both weak spin coupling ($J_0=0.072/T$, left panels) and strong spin-coupling ($J_0=0.2/T$, right panels) and different ranges of spin interactions (characterized by $\beta$ as specified in the legends). The time period is fixed by setting the drive parameters $h,\omega$ to the CDT/DL point, \textit{i.e.} the first root of $\mathcal{J}_0(4h/\omega)$. All other parameters are the same as the simulations visualized in figure~\ref{Fig:strong_weak_ea}.}
	\label{Fig:regiogionalmag}
\end{figure}
In panels `d', `e', `f' of figure~\ref{Fig:reg_mag_ea_eb}, we have plotted $M^z_{A/B}$ for a constant $\epsilon_{B} = 0.1$ and $\epsilon_{A}(=0.03,0.4,0.9)$, respectively. When $\epsilon_{A}=0.03~\&~ \epsilon_{B}=0.1$ (figure~\ref{Fig:reg_mag_ea_eb}-d), we observe that the DTC phase, present in both regions initially, melts simultaneously. As $\epsilon_{A}$ increases, a transition from a melting-DTC to a stable DMBL is observed in region A, as seen in figure~\ref{Fig:reg_mag_ea_eb}-f. Additionally, the stability of the DTC phase in region-B increases with an increase in $\epsilon_{A}$. Thus, a significantly small $\epsilon_{A}$ and large $\epsilon_{B}$ results in the  coexistence of DTC phase in region B and  DMBL phase in region A, manifesting a chimeralike order. For general comprehension, we have calculated the relative spin rotational error $\epsilon^r_{A,B} = \epsilon_{A}-\epsilon_{B}$ and plotted it in the $X_y$ panel, considering constant $\epsilon_{A} = 0.03$, and in the $X_x$ panel, considering constant $\epsilon_{B} = 0.1$. We observe that only at the extreme values of $\epsilon^r_{A,B}$, \textit{i.e.}, small $\epsilon^r_{A,B} \sim -1.0$ (figure~\ref{Fig:reg_mag_ea_eb}$X_y$) or  large $\epsilon^r_{A,B} \sim 1.0$ (figure~\ref{Fig:reg_mag_ea_eb}$X_x$), stable chimeralike states have been seen. As $\epsilon^r_{A,B}\rightarrow 0$, the chimeralike state vanishes. For tiny $\epsilon^r_{A,B}$, we observe the DTC phase in region A, and the DMBL in region B. However, for very large values of $\epsilon^r_{A,B}$, we observe the opposite: DMBL in region A and DTC in region B. Therefore, the selection of rotational errors is a crucial element in engineering and stabilizing a DTC-DMBL chimeralike state.

We extended the numerical simulations outlined in earlier sections to encompass longer durations, up to $2000T$. Consequently, we acquired long-time estimates of the regional magnetization under varying conditions of spin-coupling strength and range. The results are plotted in figure~\ref{Fig:regiogionalmag}. In both cases, the drive parameters were set to the CDT/DL point. In region B, the value of $M^Z_B$ remains constant at unity across all time for all ranges. Contrastingly, the behavior of $M^Z_A$ in region A is distinct. When the coupling is weak, $M^Z_A$ mostly persists only for all-to-all interactions. However, for interactions in other ranges, $M^Z_A$ gradually dissipates with time. When the coupling is strong, the DTC phase in region A dissolves gradually over time, regardless of the range of spin interactions. 

The melting of the DTC phase in $M^Z_A$ manifests as beats, similar to the melting of DTC observed in other systems, such as Stark-MBL~\cite{Liu2023} and integrable DTCs~\cite{Chandra2024}.
\begin{figure}[t]
	\centering
	\hspace{1.5cm}\includegraphics[width = 13cm]{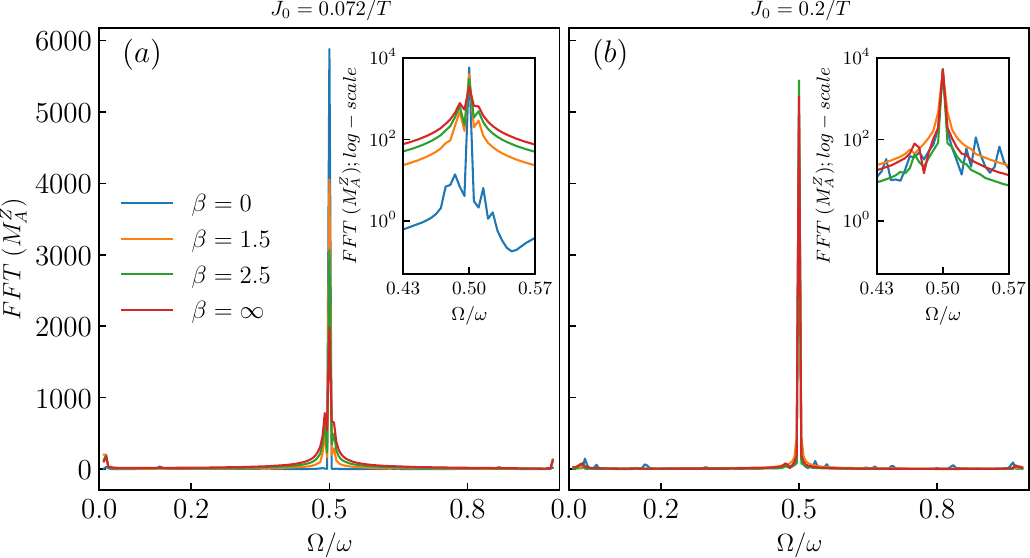}
	\caption{
		Fast Fourier Transforms(FFT) of regional magnetization $M^z_{A}$ of the spin-chain. The magnetization of regions A are plotted as functions of time ($\Omega/\omega$) for both weak spin-coupling ($J_0=0.072/T$, left panel) and strong spin-coupling ($J_0=0.2/T$, right panels) and different ranges of spin interactions (characterized by $\beta$ as specified in the legends). The time period is fixed by setting drive parameters $h,\omega$ to the first CDT/DL point, \textit{i.e.} the first root of $\mathcal{J}_0(\chi)$. All other parameters are the same as the simulations visualized in figure~\ref{Fig:strong_weak_ea}. In the inset, the variation in FFT is plotted on a logarithmic scale for a small frequency window.}
	\label{Fig:regionalFFT}
\end{figure}
The underlying mechanism can be investigated spectrally by performing FFT on the regional magnetization time-series data in a manner similar to the FFTs discussed in section~\ref{sec:level3}, except with larger datasets, and, consequently, better resolved beats. The results are depicted in figure~\ref{Fig:regionalFFT}, where the FFTs were performed on the time-series data shown in figure~\ref{Fig:regiogionalmag}. When the coupling is weak, the subharmonic peak at $\Omega=\omega/2$ are prominent in frequency space $\Omega$; the secondary peaks are barely noticeable. Consequently, sustained regional magnetization is seen over an extended period of time for all-to-all interactions. As the parameter $\beta$ is strengthened,  the subharmonic peak becomes less prominent, contributing to faster disintegration of $M^Z_A$. When the coupling is strong, the dominant peak is at $\Omega = \omega/2$, however the peaks are much less prominent across all ranges, leading to a more rapid disintegration of the regional magnetization compared to the weak spin-coupling. Therefore, a stable DTC-DMBL chimeralike state persists over a long period, with weak coupling considered in an all-to-all interaction.

\subsection{\label{sec:level41} Entanglement Entropy}

In contrast to the preceding subsection, which focused on the examination of macroscopic observables, the current section delves into a more direct analysis of microscopic aspects through the consideration of entanglement entropy (EE). The EE is commonly employed as a measure to quantify the degree of entanglement exhibited by quantum states across invariant subspaces. It provides valuable insight into quantum correlations and the ability to store information. The mathematical expression for the entanglement entropy, denoted $S_{AB}$, between areas A and B, can be described using the von-Neumann entropy~\cite{bayat_entanglement_2022,mendes-santos_measuring_2020}.   The density matrix of a pure state $\ket{\psi}$ is $\hat{\rho}= \ketbra{\psi}$. The reduced density matrices (RDMs) for regions A and B, denoted as $\hat{\rho}_A$ and $\hat{\rho}_B$ respectively, are obtained by taking the partial traces of $\hat{\rho}$ with respect to the complementary regions. Specifically, $\hat{\rho}_A$ is obtained by tracing out region B, \textit{i.e.}, $\hat{\rho}_A\equiv \Tr_B(\hat{\rho})$, while $\hat{\rho}_B\equiv\Tr_A(\hat{\rho})$. The entanglement entropy (EE) is determined by
\begin{equation} 
    S_{AB} = -\Tr\big[\hat{\rho}_A \ln\left(\hat{\rho}_A\right)\big] = -\Tr\big[\hat{\rho}_B \ln\left(\hat{\rho}_B\right)\big].
    \label{eq:vonentrop}
\end{equation}	
\begin{figure}[t!]
    \begin{center}
        \includegraphics[width=10cm]{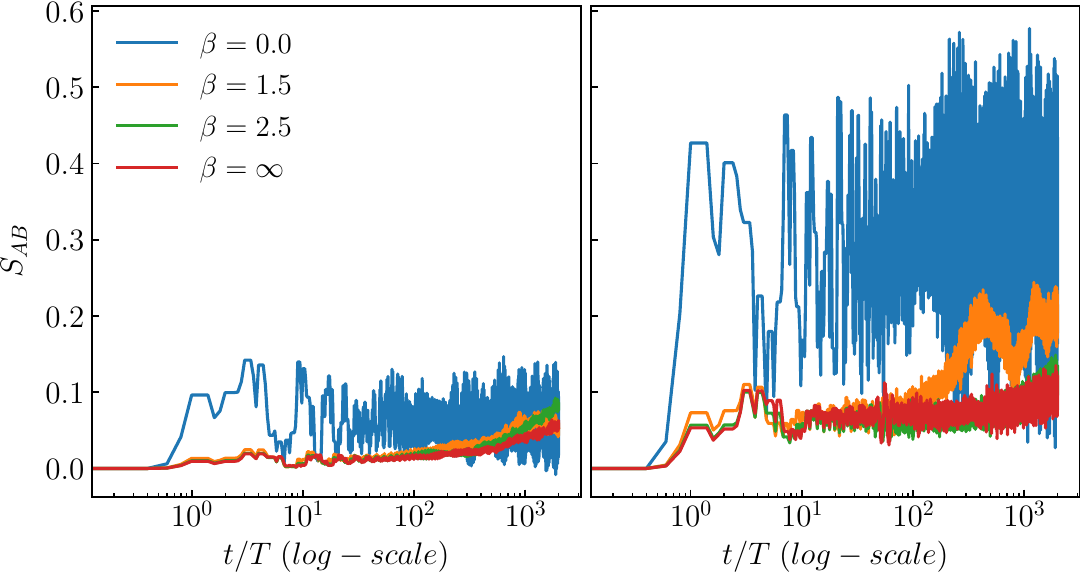}
    \end{center}
    \caption{Time evolution of Entanglement Entropy $S_{AB}$ between regions A and B as a function of time at the CDT/DL point. The left panel plot results for weak spin coupling, and the right panel plot results for stronger spin couplings. The corresponding values of $J_0$ are the same as those of figure~\ref{Fig:regiogionalmag}, and all other parameters are identical to previous simulations. The spin-spin coupling interaction in $\hat{H}_2$ during $T_2$ leads to the onset of an increase in EE. The growth rate of EE remains notably slow, persisting even up to $t=2000\;T$.}
    \label{Fig:entangle}
\end{figure}
We have extended the previous simulations to  a later time point of $t=2000\;T$, and calculated the EE for  each time point. The results are plotted in figure~\ref{Fig:entangle} for both weak and strong spin coupling, while maintaining the driving parameters consistently at a CDT/DL point. The EE is found to start to rise from the onset of the first $T_2-cycle$, when spin interactions begin to impact the behavior of the spin chain. The increase in EE is observed to be gradual, even after prolonged interactions.  The EE exhibits a more rapid increase in the case of strong spin coupling compared to weak spin coupling. As evidenced by equation~\eqref{eq:floq_eff3}, the coupling present in the effective Hamiltonian is directly proportional to the rotational error $\epsilon_A$. When the value of $\epsilon_A$ is small, it effectively hinders the interaction between regions A and B, therefore suppressing the increase in EE. This corroborates the results depicted in figure~\ref{Fig:ea}, where the DTC appears to be more stable at smaller $\epsilon_A$ than at larger $\epsilon_A$.

When thermalized at infinite temperature, the EE is extensive and averages at $\displaystyle \overline{\expval{S_{AB}}}_T\rightarrow \left[N \ln{2}-1\right]/2$ \cite{Lu2021}. When localized, the EE per particle vanishes in the thermodynamic limit. For finite sizes, it can be approximated as $\overline{\expval{S_{AB}}}_L\approx \ln{2}$~\cite{sakurai_phys_nodate}. In our spin chain, $N=8$, yielding theoretical expressions, $\overline{\expval{S_{AB}}}_T\approx 2.27$, and $\overline{\expval{S_{AB}}}_L\approx 0.69$. The numerical values of the entanglement entropy (EE) are found to remain approximately in the range of $\in\left[0,0.6\right)$, as can be seen in figure~\ref{Fig:entangle}. This observation indicates that the entire system remains athermal and localized, even when a significant amount of time has elapsed.
	
\subsection{\label{sec:level43} System size dependent stability in chimeralike state}
\begin{figure}[t!]
	\begin{center}
		\includegraphics[width=12cm]{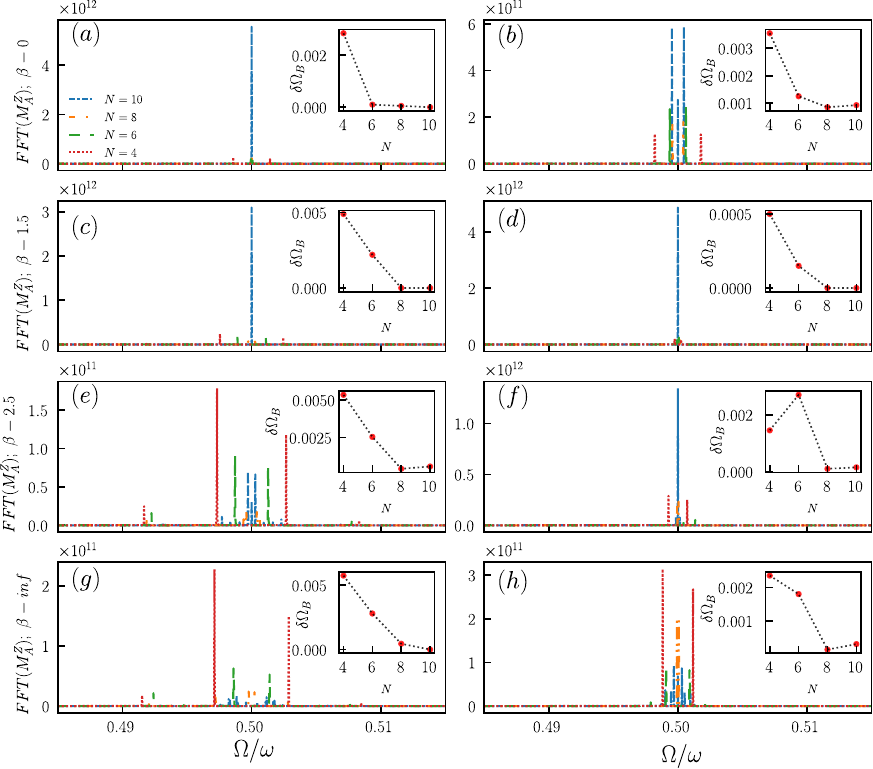}
	\end{center}
	\caption{
		FFT of regional magnetization $M^z_A$ obtained for $8\times10^4$T for different spin interaction ranges ($\beta = 0,1.5,2.5, \infty[inf]$) at panels from top to bottom, respectively. Weak (left panels) and strong (right panels) spin couplings are considered. The drive parameters are at the CDT/DL point, considering constant $\omega = 20$ with a time period of $T= 2\pi/\omega$. The beat frequency $\delta\Omega_B$ is plotted in the inset figure of each panel for different system sizes N.
	}
	\label{Fig:size_dependence}
\end{figure}	
We have observed in section~\ref{sec:level42} that DTC phase melts over time for all possible spin interactions in strong coupling as well as for all selected short-range interactions ($\beta > 0$) in weak coupling in figure~\ref{Fig:regiogionalmag}. This necessitates an investigation into the DTC phase's stability across a wide range of factors, including different system sizes. The phenomenon of DTC-melting is observed in the temporal evolution curves of $M^z_A$ in a system of size N = 8, coinciding with the occurrence of beats.

We investigated the temporal evolution of $M^z_A$ for different system sizes (N = 4, 6, 8, and 10) over an extended time period of up to $8\times10^4$T. We performed a numerical Fast Fourier Transform (FFT) on the extended time series data. The results are displayed in figure~\ref{Fig:size_dependence} for various spin interaction ranges and coupling strengths. The most prominent peaks indicate frequencies that contribute most to the system dynamics over time. We calculated the beat frequency ($\delta\Omega_B$) 
as the difference between the most prominent peaks on either side of the subharmonic peak at $\Omega/\omega = 1/2$, in a manner similar to~\cite{Chandra2024}. These were computed for different system sizes (N = 4, 6, 8, 10). Their dependence on system size are shown in the inset of each panel. As the system size increases, the change in $\delta\Omega_B$ diminishes, vanishing in the thermodynamic limit ($\lim_{N\to\infty}\delta\Omega_B= 0$),
the peaks on either side of the subharmonic merge into a single peak at $\Omega/\omega = 1/2$. This confirms that these beats are a finite-size effect, and the DTC phase is stable in the thermodynamic limit, regardless of the interaction ranges and coupling strengths.

\subsection{\label{sec:level44} Robustness against Static Fields}
\begin{figure}[t!]
	\begin{center}
		\includegraphics[width=13cm]{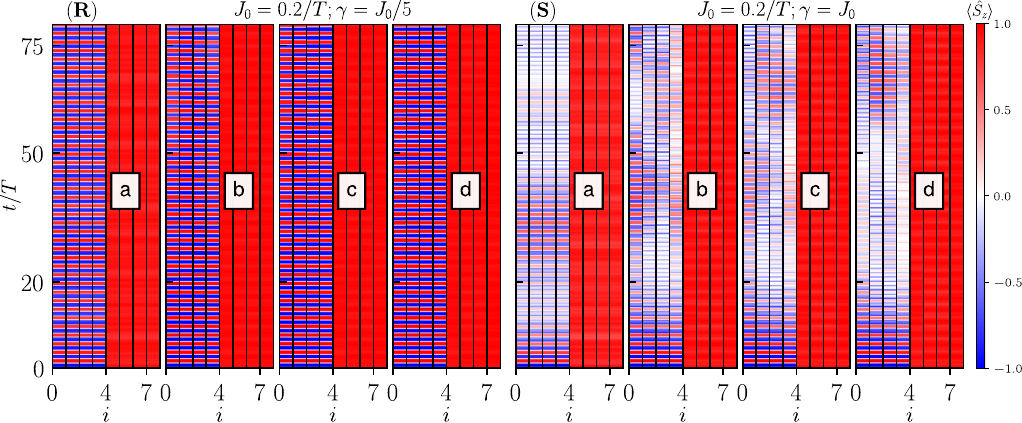}
	\end{center}
	\caption{Weak (left R- panel) and strong (right S-panel) additional static fields defined by $\gamma = J_0/5, J_0$ respectively are applied in addition to the strong ($J_0=0.2/T$) spin coupling and transverse field at different interaction ranges set by $\beta = 0$ (panels a), $\beta= 1.5$ (panels b), $\beta=2.5$ (panels c), and $\beta= \infty$ (panels d). The frequency of the drive is $\omega = 20$, amplitude $h$ is corresponding to CDT/DL point, $\epsilon_A=0.03$, $\epsilon_B=0.9$.  The local magnetization for each site (i) is plotted from $t=0$ to $t=80 T$.}
	\label{Fig:robustness}
\end{figure}
The choice of the driving parameters $h, \omega$ significantly influences the persistence of the stability of the DTC-DMBL chimeralike state, especially when they deviate from the CDT/DL point. When there is a small deviation , the chimera-like order stays stable (see ~\ref{sec:AppendixC} for details). However, the system becomes less stable for larger deviations, as seen in figure~\ref{Fig:strong_weak_ea}. Since the inclusion of weak static fields to a system at CDT/DL point can be qualitatively viewed as deviations of the clean system from the CDT/DL point, we can infer that static fields will notably influence the stability of the DTC-DMBL chimeralike state. In order to conduct an investigation of this influence, we incorporated the static field term $\hbar\hat{V}(\hat{\sigma}^{\gamma})$ (as discussed in section~\ref{sec:mdl_n_dynam}) into the Hamiltonian $\hat{H}(t)$. This static field remains unmodulated by the pulses and acts on the system identically at all times. We have conducted the simulation incorporating this field while all other parameters are maintained at the same values at CDT/DL point as in earlier simulations. 

We have run simulations for both strong ($\gamma=0.2/T$) and weak ($ \gamma= 0.2/5T$) field amplitudes. The findings are displayed in figure~\ref{Fig:robustness}. At low values of $\gamma$, the influence of spin interactions is more noticeable than the influence of the static field, leading to the continued existence of dynamical localization and a stable chimeralike state of the DTC. However, when the value of $\gamma$ increases, the static field dominates over the interactions, resulting in the melting of the DTC after a few cycles. However, it is worth noting that the DMBL phase continues to exist even in the presence of the static field in a manner similar to DMBL reported in the literature~\cite{Mahbub2024}. Therefore, a chimeralike phase known as \textit{Thermal-DMBL} becomes apparent at longer time scales. Therefore, it is possible to achieve a strong DTC-DMBL chimeralike state even when moderate external fields are present.
	
\section{\label{sec:level7} Summary and Conclusion}
In this work, we have proposed a novel quantum chimeralike state, where a discrete-time crystal (DTC) and a dynamically many-body localized (DMBL) phase coexist in a one-dimensional spin-1/2 chain. We have divided the lattice into two regions, labeled `A' and `B', based on their rotational errors ($\epsilon_{A/B}$). We have applied two different Hamiltonians, modulated by a time-periodic square wave, which acts on the spin chain at different times. One Hamiltonian conducts spin-flip operations only in region `A' during the first duty cycle of the square wave (for time $T_1$), while the other Hamiltonian induces global dynamical localization in the entire spin chain at the second duty cycle (for time $T_2$). This process repeats at every time-period of the square wave, resulting in the coexistence of a DTC phase in region A and a DMBL phase in region B. We have applied specific high-frequency periodic drive parameters engineered to attain dynamical localization via Coherent Destruction of Tunneling (the CDT/DL point) during the $T_2$ cycle. We have produced the results of numerical simulations that have shown that the interaction on the spin chain plays an essential role in the formulation of the chimeralike state and the stabilization of the DTC phase. Our results suggest that the DTC does not exhibit temporal persistence when the spin-spin coupling strength is weak. This was seen for all types of interaction ranges (except for all-to-all interactions). The presence of a strong spin coupling hinders the relaxation of spins away from the DTC, stabilizing the DTC phase regionally for long times throughout all interaction ranges. Interestingly, the stability of the DTC in region A is contingent on the stability of the DMBL phase in region B. The weak coupling with long-range spin interaction is very robust against the exchange energy per spin, and this enables the chimeralike state to persist for long time. Analytically, the inter-region coupling term in the effective Floquet Hamiltonian in equation~\ref{eq:floq_eff3} is directly proportional to the spin rotation error $\epsilon_A$. We have tuned $\epsilon_A$ and found that the chimeralike state can persist only if $\epsilon_A$ is small. In the case of larger values, there is an increased level of coupling between the two regions. This heightened coupling results in the immediate melting of the DTC for all interactions, except for long-range interactions.
	
To further study the resilience of this chimeralike state against static perturbations, an additional external static field is included. Numerical simulations have demonstrated that the system is robust in the face of minor perturbations, as evidenced by the sustained presence of the DTC phase inside region A for an extended duration. Larger perturbations destabilize the system, dissolving the DTC. Analytically, the Floquet Hamiltonian, obtained in the RWA, illustrates the coupling between regions A and B. The entanglement entropy rises at a very slow rate for both weak and strong regional spin interactions. Therefore, the entire system is effectively prevented from thermalizing. Additionally, the persistence of the DTC phase is directly dependent on the stability of the DMBL phase. Our proposed DTC-DMBL chimeralike model shows sustainability even at larger times. In the thermodynamic limit, our proposed model is found to be stable for all types of spin interactions and spin coupling strengths. 

Such chimeralike states can be experimentally realized in systems where DMBL has already been seen. For example, a quantum spin-$1/2$ system consisting of $^{19}\mathrm{F}$ trifluoroiodoethylene in acetone-D6~\cite{Hegde2014}, as well as trapped ions~\cite{sakurai_phys_nodate, Friedenauer2008}, as described in previous studies. However, there are limits to the resistance to external fields, even while functioning precisely at the CDT/DL point in their absence.

Such chimeralike states will be beneficial for quantum engineering and quantum memory design based on DTC~\cite{zhang_observation_2017}. Furthermore, our model may provide new insights into the design of quantum neural networks, such as quantum reservoir computing systems~\cite{Fujii_2017, Martinez_2021,Mujal_2021, Akitada2022}.  Additionally, in any chimera state, the part of the system undergoing incoherent dynamics can be viewed as a reservoir that is coupled to the coherently-oscillating system. This coupling must be non-Markovian in nature, since information that is lost to the environment can potentially return to the system~\cite{Accardi1976,Lindblad1976}. Chimeralike states are therefore useful for studying these types of non-Markovian dynamics in quantum many body physics, extending our understanding of open systems beyond the conventional Markov limit,  where information is permanently lost to the environment~\cite{Agarwal2012, Gorini1976}. Finally, chimera states have potential applications in modern quantum devices. They can be used in NISQ processors to distinguish between the relatively clean and noisy qubits. Normally, information diffusion starts with the noisy qubits, leading to the quantum state slowly losing coherence into a classical mixed state~\cite{Bharti2022}. However, using a chimeralike state, the clean qubits can be shielded, effectively extending the coherence time. Thus, our model enables more sophisticated quantum information processing. 
	
\ack{MR acknowledges The University of Burdwan for support through the State-Funded fellowship. AR acknowledges support from the University Grants Commission (UGC) of India, via BSR Startup Grant No. F.30-425/2018(BSR), also from the Science and Engineering Research Board (SERB, Govt. of India) Core Research Grant No. CRG/2018/004002. Additionally, AR thanks Prof. Tanmoy Banerjee, Dept. of Physics, The University of Burdwan, for useful discussions.}
\medskip

\clearpage

\appendix
\section{\label{sec:AppendixA} DMBL in $T_2$ time interval}

Let us consider a one-dimensional spin-$1/2$ chain consisting of $N$ spins, with Heisenberg exchange interaction between them. The interaction follows a power law decay rule, where the exchange energy between the $i^{th}$ and $j^{th}$ spins is presumed to be  $J_{ij}=J_0\abs{i-j}^{-\beta}$. Let us now  divide the chain into two regions, A and B by manipulating corresponding spin rotational errors, and introduce time dependencies via two pulse wave sequences in the manner described in section~\ref{sec:mdl_n_dynam} and figure~\ref{Fig:time_distribution}. The first sequence  induces spin-flips in region A within time $T_1$, and the second modulates a global sinusoidal periodic drive in the field \textit{s.t.} $h_D = -h\sin(\omega t)$ within time $T_2$, chosen to be such that the time period $T\equiv 2\pi/\omega = T_1 + T_2$ and $T_1 = T_2$. This drive parameters are controlled in such a way that the system is dynamically localized during the $T_2-$cycles.

The Hamiltonian during the $T_2-$cycle is $\displaystyle \hat{H}_2(t) = \hbar\sum_{i>j} J_{ij}\hat{\sigma}^y_i\hat{\sigma}^y_j - \hbar h \sin(\omega t) \sum_i \hat{\sigma}^z_i$. Let us define the free propagator 
\begin{equation}
    \hat{U}(t) \equiv \exp \bigg[-\frac{i}{\hbar}\int_{\frac{T}{2}}^{\frac{T}{2}+t} (-h \sin(\omega t'))dt' \hbar\sum_i\hat{\sigma}^z_i\bigg]
    =\prod_{i} \exp\big[-i \hat{\sigma}^z_i\zeta(t)\big],
\end{equation}
where, $\displaystyle \zeta (t) = h\int_{\frac{T}{2}}^{\frac{T}{2}+t}  (-\sin(\omega t')dt') =\frac{h}{\omega}(1-\cos(\omega t))$.
Now, the Hamiltonian can be transformed to the moving frame as follows \cite{haldar_statistical_2022}.
\begin{align}
    \hat{H^{mov}}(t) &= \hat{U}^\dagger(t) \hat{H}_2(t) \hat{U}(t)- i \hat{U}^\dagger(t) \partial_t \hat{U}(t) \nonumber\\
    &= \prod_{i} \exp\big[-i\hat{\sigma}^z_i \zeta(t)\big] \big[\hbar\sum_{ij}J_{ij}\hat{\sigma}^y_i\hat{\sigma}^y_j\big] \exp\big[ i\hat{\sigma}^z_i \zeta(t)\big]\nonumber\\
    &= \hbar\sum_{ij} J_{ij}\;\big[ e^{-i\hat{\sigma}^z_i \zeta(t)} \hat{\sigma}^y_i e^{i\hat{\sigma}^z_i \zeta(t)}\big]\quad \big[e^{-i\hat{\sigma}^z_j  \zeta(t)}\hat{\sigma}^y_j e^{i\hat{\sigma}^z_j \zeta(t)}\big]\nonumber\\
    &=\hbar\sum_{ij}J_{ij}\Big\{\hat{\sigma}^y_i\cos{\big[2\zeta(t)\big]}-\hat{\sigma}^x_i\sin{\big[2\zeta(t)\big]}\Big\}\Big\{\hat{\sigma}^y_j\cos{\big[2\zeta(t)\big]}-\hat{\sigma}^x_j\sin{\big[2\zeta(t)\big]}\Big\},\nonumber\\
    &= \hbar\sum_{ij} J_{ij}  \Big[ \hat{\sigma}^y_i\hat{\sigma}^y_j\cos^2(2\zeta) +\hat{\sigma}^x_i\hat{\sigma}^x_j \sin^2(2\zeta) - \frac12 (\sigma^y_i\hat{\sigma}^x_j + \hat{\sigma}^x_i\hat{\sigma}^y_j)\sin(4\zeta)\Big].
    \label{eq:hmovap1}
\end{align}
In the penultimate step, we used the Baker-Campbell-Hausdorff formula~\cite{Magnus1954}. Now, we invoke the Jacobi-Anger expansion, $e^{i\eta \cos(\theta)} = \sum_{n=-\infty}^{\infty}\mathcal{J}_n(\eta)e^{in\theta}$, where $\mathcal{J}_n(z)$ is the Bessel function of the first kind of $n^{th}$ order, and apply to the RHS of  equation~\ref{eq:hmovap1}. This produces a Fourier series expansion for each amplitude in
$\hat{H}^{mov}(t)$ with period $\omega$. In high frequency limit, where $\omega \gg J_0$, the Rotating Wave Approximation (RWA) allows for approximating the Fourier modes with their long time averages, thus removing the contributions of all fast-oscillating terms. This yields approximations
\begin{align}
\cos[2](2\zeta) &\approx \frac12 \Bigg[1+ \cos(\frac{4h}{\omega})\mathcal{J}_0\left(\frac{4h}{\omega}\right)\Bigg]\nonumber\\
\sin[2](2\zeta) &\approx \frac12 \Bigg[1- \cos(\frac{4h}{\omega})\mathcal{J}_0\left(\frac{4h}{\omega}\right)\Bigg]\nonumber\\
\sin(4\zeta)&\approx \frac12 \sin(\frac{4h}{\omega})\mathcal{J}_0\left(\frac{4h}{\omega}\right).
\label{eq:sqr}
\end{align}
Finally, we substitute the values from equation~\eqref{eq:sqr} into equation~\eqref{eq:hmovap1}. This reduces the moving frame Hamiltonian into a RWA-Hamiltonian ($\hat{H^{_{RWA}}}$) given below,
\begin{multline}
    \hat{H}^{mov}\approx \hat{H}^{_{RWA}} = \frac{\hbar}{2}\sum_{ij} J_{ij}  \Bigg\{ \hat{\sigma}^y_i\hat{\sigma}^y_j\Bigg[1+ \cos(\frac{4h}{\omega})\mathcal{J}_0\left(\frac{4h}{\omega}\right)\Bigg] +\hat{\sigma}^x_i\hat{\sigma}^x_j \Bigg[1- \cos(\frac{4h}{\omega})\mathcal{J}_0\left(\frac{4h}{\omega}\right)\Bigg] \\
    - \frac12 (\sigma^y_i\hat{\sigma}^x_j + \hat{\sigma}^x_i\hat{\sigma}^y_j)\sin(\frac{4h}{\omega})\mathcal{J}_0\left(\frac{4h}{\omega}\right)\Bigg\}.
    \label{eq:hrwa}
\end{multline}
Now, for a particular $\omega$, if $h$ is controlled in such fashion that $\frac{4h}{\omega}$ lies on the CDT/DL point, given by a root of $\mathcal{J}_0\Big(\frac{4h}{\omega}\Big)$, then $\hat{H}^{_{RWA}}$ becomes the "CDT/DL Hamiltonian",
\begin{equation}
\hat{H}^{_{RWA}}_{_{CDT/DL}} = \frac{\hbar}{2}\sum_{ij} J_{ij}  \Big[\hat{\sigma}^x_i\hat{\sigma}^x_j + \hat{\sigma}^y_i\hat{\sigma}^y_j\Big].
\label{eq:hrwa:frz}
\end{equation}	
One limiting case of the $J_{ij}$s is the nearest-neighbor case, where $J_{ij} = J_0 \delta_{i\;j\pm1}$. In that case, performing a Jordan-Wigner Transformation ,
\begin{align}
\hat{\sigma}^x_i &= \left(\hat{c}_i + \hat{c}^\dagger_i\right) \prod_{j=0}^{i-1}\left(1-2\hat{c}^\dagger_j \hat{c}^{\;}_j\right),\nonumber\\
\hat{\sigma}^y_i &= i\left(\hat{c}_i - \hat{c}^\dagger_i\right) \prod_{j=0}^{i-1}\left(1-2\hat{c}^\dagger_j \hat{c}^{\;}_j\right),\nonumber\\
\hat{\sigma}^z_i &= 1-2\hat{c}^\dagger_i\hat{c}^{\;}_i,
\end{align}
maintains the locality of the Hamiltonian in the basis of fermion creation and annihilation operators $\hat{c}^{\;}_i, \hat{c}^\dagger_i$ \cite{Lieb1961, mbeng2020}. Here, the CDT/DL Hamiltonian transforms to the standard tight-binding Hamiltonian $\hat{H}^{_{RWA}}_{_{CDT/DL}}=(J_0/2)\sum_i \hat{c}^\dagger_i \hat{c}^{\;}_{i+1}$. The system is completely localized due to the presence of $N$ local (approximate) Noether Charges, given by $\hat{n}_m= \hat{d}^{\dagger}_m\hat{d}^{\;}_m$, where
\begin{equation}
\hat{d}_m = \frac{1}{\sqrt{N}}\sum_{n=0}^{N-1}e^{-\frac{2nm\pi i}{N}}\hat{c}_n.
\end{equation}
When the $J_{ij}$s are extended beyond the nearest-neighbor case to longer ranges, the locality of the Jordan Wigner transformation is destroyed. Nonetheless, the normal dynamics of the transverse field remains frozen at the CDT/DL point , since $\comm{H^{_{RWA}}_{_{CDT/DL}}}{\displaystyle\sum_i\sigma^z_i}=0$.

In our driving protocol, the system is always ideally populated in an eigenstate of the transverse field at the end of every $T_1$ time interval, with a $z-$polarized region A and a region B that is polarized in the opposite direction. Thus, the field dynamics will, in the ideal case, be frozen in all $T_2$ time intervals.

\section{\label{sec:AppendixB} Effective Floquet Hamiltonian}

Floquet theory is a widely used methodology for evaluating the dynamical behavior of time-periodic systems.  In a quantum system that is time periodic with period $T$, the Hamiltonian obeys $\hat{H}(t+T) = \hat{H}(t)\;\forall t$. If we split the time-dependent part from the time-independent, or \textit{d.c.} part, we can write
\begin{equation*}
    \hat{H}(t) = \hat{H}_0 + \varepsilon \hat{H}_1(t)
\end{equation*}
The corresponding propagator can be obtained by solving the Schr\"odinger equation $\displaystyle{i\hbar \partial_t \hat{U}(t) = \hat{H}(t) \hat{U}(t)}$.  The propagator at $t=T$, given by $ \hat{\mathcal{F}}\equiv \hat{U}(T)$ is called the \textit{Floquet operator}. Now, it follows from Floquet's Theorem that, if the system is strobed at integer multiples of $T$, the dynamics can be mapped to that of a time-independent effective Hamiltonian $\hat{H}_F$, such that  $\displaystyle\hat{U}(nT) = \exp\bigg[-\frac{i}{\hbar}\hat{H}_F\; nT \bigg]$~\cite{Eckardt_2015}.  The operator$\hat{H}_F$ (also denoted by $H^\mathrm{eff}$) is obtained from the original time-dependent Hamiltonian, and is given by, $\displaystyle \hat{H}_F = \Bigg(\hat{H}(t) - i\hbar \pdv{t}\Bigg)$ evaluated at $t=T$.

Now, the proposed spin chain in equation~\eqref{eq:cleanham} needs to evolve for at least two time periods to manifest a DTC. If the effective time-independent Hamiltonian at that time is denoted by $\hat{H}^{\mathrm{eff}}_{\epsilon_A, 2T}$, then, the propagator
\begin{align}
    \hat{\mathcal{F}}^{2} &\equiv \exp\Big(-\frac{i}{\hbar}\hat{H}_{\epsilon_A, 2T}^{\mathrm{eff}}\;2T\Big) \label{eq:heff}\\ 
    &= \mathcal{T}\exp\Big(-\frac{i}{\hbar}\int_{\frac{3T}{2}}^{2T}\hat{H}_2(t) dt\Big)
    \exp\Big(-\frac{i}{\hbar}\hat{H}_1T_1\Big)\nonumber\\
    & \hskip 5cm \mathcal{T}\exp\Big(-\frac{i}{\hbar}\int_{\frac{T}{2}}^{T}\hat{H}_2(t) dt\Big)\exp\Big(-\frac{i}{\hbar}\hat{H}_1T_1\Big).
    \label{eq:2tprop}
\end{align}
Here $\mathcal{T}$ denotes the time-ordering operation. 
To evaluate $\hat{\mathcal{F}}^2$, we start by calculating each of the four terms in RHS of equation~\ref{eq:2tprop}. During the $T_1-$cycle, $t\in[0,T/2]$, and the time-independent Hamiltonian $\hat{H}_1$ is applied on spin chain. Thus,
\begin{equation}
    \exp\Big(-\frac{i}{\hbar} \hat{H}_1 T_1\Big) = 	\exp\Bigg[\frac{-i(1-\epsilon_A)\pi}{2}\sum_{l \in A}\hat{\sigma}^x_l\Bigg].
    \label{eq:effham1}
\end{equation}	
It is evident that the effective Hamiltonian for $t \in[T, 3T/2]$ will be same as that obtained from equation~\eqref{eq:effham1}. Next, during the $T_2-$cycle, $t\in[T/2,T]$, and the time-dependent Hamiltonian, $\hat{H_2}(t)$ is applied. Therefore,
\begin{align}
    \mathcal{T}\exp\Big(-\frac{i}{\hbar}\int_{\frac{T}{2}}^{T}\hat{H}_2(t) dt\Big) &= \exp( -\frac{i}{\hbar}\int_{\frac{T}{2}}^{T}\Big[\hbar\sum_{i\neq j}J_{ij}\hat{\sigma}^y_i\hat{\sigma}^y_j-\hbar h\sin(\omega t)\sum_i\hat{\sigma}^z_i\Big] dt)\nonumber\\
    &= \exp(-i \Big[\sum_{i\neq j}J_{ij}\hat{\sigma}^y_i\hat{\sigma}^y_j \frac{T}{2} + \frac{2h}{\omega}\sum_i \sigma^z_i\Big])
    \label{eq:t2hamilt}
\end{align}
Clearly, the effective Hamiltonian for the times  $t \in[3T/2, 2T]$ will also be the same as equation~\eqref{eq:t2hamilt}. Now,	let us consider the case where $\epsilon_A \neq 0$ and $\epsilon_B=1$. Furthermore, let us define the operators $\displaystyle \hat{V}_{\epsilon_A} \equiv \exp\Big(\frac{i\epsilon_A \pi}{2}\sum_{l\in A}\hat{\sigma}^x_l\Big), \displaystyle \hat{\theta}_l \equiv 2 \left(\sum_{m \in B}J_{lm}\hat{\sigma}^y_m \frac{T}{2} \right)$; the latter describes a rotation acting on operator $\hat{\sigma}^y_l$ in only region B.  Now, we can expand expression for the  $T_2-$cycle propagator in equation~\ref{eq:t2hamilt} by splitting the contributions from regions A and B. This yields
\begin{multline}
    \mathcal{T}\exp\bigg\{-\frac{i}{\hbar}\int_{\frac{T}{2}}^{T} \hat{H}_2(t) dt\bigg\} = \exp\Bigg\{-i \bigg[ \sum_{l,m\in A}J_{lm} \hat{\sigma}_l^y\hat{\sigma}_m^y +\sum_{l,m\in B}J_{lm}
    \hat{\sigma_l^y}\hat{\sigma_m^y}+\sum_{\substack{%
            l \in A,\\
            m \in B\hfill}} J_{lm}\hat{\sigma}^y_l\hat{\sigma}^y_m\\
    +\frac{4h}{\omega T}\big( \sum_{l\in A}^{}\hat{\sigma}^z_l + \sum_{m\in B}^{}\hat{\sigma}^z_m\big)\bigg]\frac{T}{2}\Bigg\}.
    \label{eq:effham3}
\end{multline}	
Substituting the RHS of equations~\eqref{eq:effham1} \& \eqref{eq:effham3} into the RHS of equation~\eqref{eq:2tprop} yields    	
\begin{multline}
    \hat{\mathcal{F}}^2 
    = \exp\Bigg\{-i \Bigg[ \sum_{l,m\in A}J_{lm} \hat{\sigma}^y_l\hat{\sigma}^y_m\frac{T}{2} +\sum_{l,m\in B}J_{lm} \hat{\sigma}^y_l\hat{\sigma}^y_m\frac{T}{2}+\sum_{\substack{%
            l \in A,\\
            m \in B\hfill}} J_{lm}\hat{\sigma}^y_l\hat{\sigma}^y_m\frac{T}{2} \\
            +\frac{2h}{\omega }\left( +\sum_{l\in A}^{}\hat{\sigma}^z_l + \sum_{m\in B}^{}\hat{\sigma}^z_m\right)\Bigg]\Bigg\} \hat{V}_{\epsilon_A} \exp\Bigg\{-i \Bigg[ \sum_{l,m\in A}J_{lm} \hat{\sigma}^y_l\hat{\sigma}^y_m\frac{T}{2}\\ +\sum_{l,m\in B}J_{lm} \hat{\sigma}^y_l\hat{\sigma}^y_m\frac{T}{2}
            -\sum_{\substack{%
            l \in A,\\
            m \in B\hfill}} J_{lm}\hat{\sigma}^y_l\hat{\sigma}^y_m\frac{T}{2} +\frac{2h}{\omega}\left( -\sum_{l\in A}^{}\hat{\sigma}^z_l + \sum_{m\in B}^{}\hat{\sigma}^z_m\right)\Bigg]\Bigg\} \hat{V}_{\epsilon_A}.\nonumber
\end{multline}
We can simplify this expression further with the Suzuki–Trotter decomposition\cite{Ostmeyer_2023, Hatano2005} that all operators with norms $\mathcal{O}(T^2)$ can be neglected in comparison to those with norm $\sim T$ for sufficiently large $\omega\equiv 2\pi/T$. Thus, for instance, $e^{T\hat{A} + T\hat{B}}= e^{T\hat{A}}\; e^{T\hat{B}}\;e^{C_2 T^2\left[\hat{A}, \hat{B}\right]}\;e^{C_3 T^3\left[\hat{A},\left[\hat{A}, \hat{B}\right]\right]}\;\dots \approx e^{T\hat{A}} e^{T\hat{B}}$, once all higher order operators are neglected after applying the Zassenhaus' formula~\cite{Magnus1954}. This yields
\begin{multline}	
    \hat{\mathcal{F}}^2 	\approx\exp\Bigg[-i \sum_{l,m\in A}J_{lm} \hat{\sigma}^y_l\hat{\sigma}^y_m\frac{T}{2}\Bigg] \exp\Bigg[-i\left(\frac{2h}{\omega } \sum_{l\in A}^{}\hat{\sigma}^z_l + \sum_{l \in A}\frac{\hat{\theta}_{l}}{2}\hat{\sigma}^y_l\right)\Bigg] 
    \hat{V}_{\epsilon_A} \\
    \exp\Bigg[i \left(\sum_{l \in A}\frac{\hat{\theta}_{l}}{2}\hat{\sigma}^y_l + \frac{2h}{\omega} \sum_{l\in A}^{}\hat{\sigma}^z_l\right)\Bigg] \exp\Bigg[-i\sum_{l,m\in A}J_{lm} \hat{\sigma}^y_l\hat{\sigma}^y_m\frac{T}{2}\Bigg] \exp\big[-i H_B T\big]\hat{V}_{\epsilon_A},\nonumber
\end{multline}
where, for the sake of brevity, we have defined the $T_2-$cycle Hamiltonian in region-B, $\displaystyle {H}_B \equiv \sum_{l,m\in B} J_{lm} \hat{\sigma}^y_l\hat{\sigma}^y_m + \frac{4h}{\omega T}\sum_{m \in B}\hat{\sigma^z_m}$. We now observe that $\hat{\theta}_l$ commutes with all operators that live in region$-A$, thus allowing us to temporarily treat it as a $c-$number in algebraic manipulations, allowing for the simplification
\begin{align}		
    \hat{\mathcal{F}}^2 	\approx& \exp\Bigg[-2i  \sum_{l,m\in A}J_{lm} \hat{\sigma}^y_l\hat{\sigma}^y_m\frac{T}{2}\Bigg]\exp\Bigg[\frac{i \epsilon_A \pi}{2}\sum_{l\in A}\Bigg\{\hat{\sigma}^x_l \cos(\hat{\theta}_l)\cos(\frac{4h}{\omega})\nonumber\\
    &+ \hat{\sigma}^y_l \cos(\hat{\theta}_l)\sin(\frac{4h}{\omega})-\hat{\sigma}^z_l \sin(\hat{\theta}_l)\Bigg\}\Bigg] \exp\big[-i H_B T\big]\hat{V}_{\epsilon_A}\nonumber\\
    =&\exp\Bigg[-2i \sum_{l,m\in A}J_{lm} \hat{\sigma}_l^y\hat{\sigma}_m^y\frac{T}{2}\Bigg] \exp\Big(\frac{i\epsilon_A \pi}{2}\sum_{l\in A}\hat{\sigma^x_l}\Big) \exp\Bigg[\frac{i \epsilon_A \pi}{2}\sum_{l\in A}\Bigg\{\hat{\sigma}^x_l \cos(\hat{\theta}_l)\nonumber\\
    & \cos(\frac{4h}{\omega})+\hat{\sigma}^y_l \cos(\hat{\theta}_l)\sin(\frac{4h}{\omega})-\hat{\sigma}^z_l \sin(\hat{\theta}_l)\Bigg\}\Bigg]\exp\big[-i H_B T\big]\nonumber\\
    =&\exp\Bigg[-2i\sum_{l,m\in A}J_{lm}\hat{\sigma}^y_l\hat{\sigma}^y_m \frac{T}{2}\Bigg] \exp\Bigg[-i\epsilon_A \pi \sum_{\substack{l,m \in A\\l\neq m}}J_{lm} \hat{\sigma}^z_l\hat{\sigma}^y_m\frac{T}{2}\Bigg] 
     \exp\Bigg[\frac{i \epsilon_A \pi}{2}\sum_{l\in A}\Bigg\{\hat{\sigma}^x_l\nonumber\\
     &\Bigg(1+ \cos(\hat{\theta}_l)\cos(\frac{4h}{\omega})\Bigg)
    +\hat{\sigma}^y_l \cos(\hat{\theta}_l)\sin(\frac{4h}{\omega})-\hat{\sigma}^z_l \sin(\hat{\theta}_l)\Bigg\}\Bigg] \exp\big[-i H_B T\big].
\end{align}
Next, we apply the Baker-Campbell-Hausdorff formula~\cite{Magnus1954} successively to all pairs of exponents in the RHS, neglecting all operators with norms $\mathcal{O}(T^2)$ as before, followed by substitution into the RHS of equation~\ref{eq:2tprop}. Finally, substitution into equation~\ref{eq:heff} yields the effective Floquet Hamiltonian in the high-frequency limit, 
\begin{multline}
    H^{\mathrm{eff}} \approx\frac{\hbar}{2} \sum_{l,m\in A}J_{lm}\hat{\sigma}_l^y\hat{\sigma}_m^y +\frac{\hbar \epsilon_A \pi}{4} \sum_{\substack{l,m\in A\\l\neq m}} J_{lm}\hat{\sigma}^z_l\hat{\sigma}^y_m + \frac{\hbar}{2}\sum_{l,m\in B}J_{lm}\hat{\sigma}_l^y \hat{\sigma}_m^y + \frac{h\hbar}{\pi}\sum_{m \in B}\hat{\sigma}^z_m \\ -\frac{\hbar \pi \epsilon_A}{4T}\sum_{l\in A}\Bigg\{\hat{\sigma}^x_l \bigg[\cos(\hat{\theta}_l)\cos(\frac{4h}{\omega})+1 \bigg] + \hat{\sigma}^y_l \cos(\hat{\theta}_l)\sin(\frac{4h}{\omega})-\hat{\sigma}^z_l \sin(\hat{\theta}_l)\Bigg\}.
    \label{eq:app:nfloq_eff3}
\end{multline}
In the limit of very high frequencies, $T\rightarrow 0$, leading to asymptotically vanishing $\hat{\theta}_l$. In this limit, the coupling between regions manifests as effective fields $\sim \hat{\sigma}^{x,y}_l$ in region $A$ whose intensities depend only on $\epsilon_A$, and the ratio $\frac{4h}{\omega}$. When $h$ is set to an odd multiple of $\pi\omega/4$, these effective fields vanish in the asymptotic limit, causing the regions $A$ and $B$ to decouple completely.

\section{\label{sec:AppendixC} Stability around CDT/DL point}
\begin{figure}[t]
	\begin{center}
		\includegraphics[width=12cm]{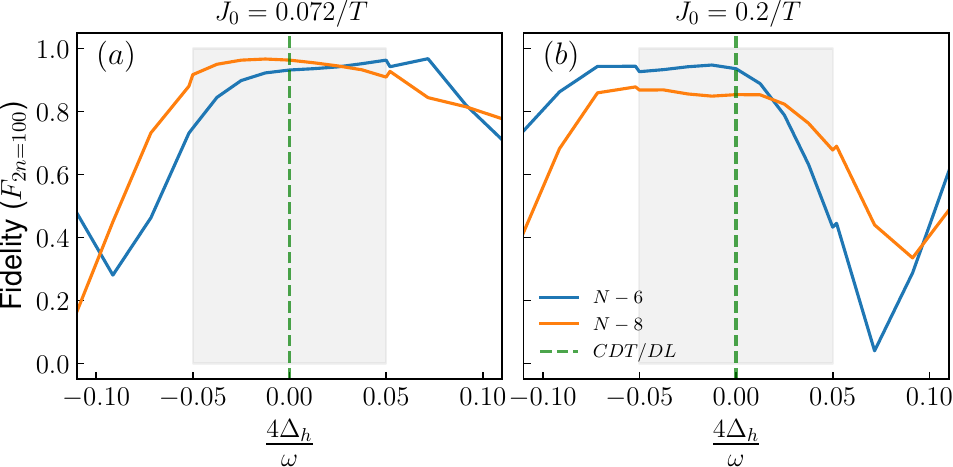}
	\end{center}
	\caption{Fidelity in relation to different deviation values, shown as $\frac{4\Delta_h}{\omega}$, for different system sizes (N) at 100T close to the CDT/DL point when spin coupling is weak (panel a) and strong (panel b). With a steady drive frequency of $\omega= 20$, the amplitude deviation ($\Delta_h$) is incorporated into h. It's observed that the fidelity is significantly high near the CDT/DL point, which is indicated by a gray-colored area.}
	\label{Fig:aroundCDT}
\end{figure}
In order to study regional DMBL, we have used the analytical results in ~\ref{sec:AppendixA} and ~\ref{sec:AppendixB}, where drive parameters were fixed at a CDT/DL point. Now, if we move away from this point, say, to a value of $6.0$ (which differs from the nearest root of $\mathcal{J}_0(4h/\omega)\approx 6.3802$ by approximately $0.40$), the stability of the chimeralike order is significantly reduced. Thus, the maximum deviation from a root must be bounded above, prompting a more detailed investigation of the stability of the chimeralike order. 

We set $\beta=0$ as shorter ranges show stronger finite-size effects (see section \ref{sec:level43}) and evolved the dynamics as described by the equation~\ref{eq:sysham2} from a fully $z-$polarized state. The drive frequency is fixed at a value ($\omega=20$) large enough for RWA to hold, and the amplitude ($h$) is set to the default CDT/DL point. We introduce a small deviation, $\Delta_h$, to $h$, and numerically calculate the fidelity ($F_{2n}$). The fidelity is a measure of how close  the time varying state is to the initial state. We have evaluated $F_{2n} = \abs{\braket{\psi(0)}{\psi (2nT)}}^2$ ~\cite{Jozsa1994,Liu2023}, where $\ket{\psi(t)}$ denotes the wave function of the system at time `$t$' and `$n=50$' is the number of double-periods chosen (\textit{i.e.}, at a time of $100T$). We have numerically obtained $F_{2n}$ for several $\Delta_h$'s and plotted them in figure~\ref{Fig:aroundCDT} for different system sizes $N = 6, 8$. The fidelity is found to nearly plateau at unity around the CDT/DL point ($\frac{4\Delta_h}{\omega} = 0$) in the slightly grayed region in the figure. 
Beyond this region, however, fidelity decreases more rapidly, manifesting a melting chimeralike state. This confirms the system's stability at and near the CDT/DL point.

\bibliographystyle{iopart-num}
\bibliography{chimera}
\end{document}